\documentclass[preprint,review,12pt]{elsarticle}
\usepackage[a4paper,margin=1in]{geometry}

\makeatletter
\def\ps@pprintTitle{%
\let\@oddhead\@empty
\let\@evenhead\@empty
\def\@oddfoot{}%
\let\@evenfoot\@oddfoot}
\makeatother
\usepackage{graphicx, subcaption}
\usepackage{lineno}
\usepackage{epstopdf}
\usepackage{color}
\usepackage{multirow}
\usepackage{textcomp,gensymb}
\usepackage{lineno}
\usepackage{xcolor}
\usepackage{physics}
\usepackage{epsfig}
\usepackage{amsmath}
\usepackage{amssymb}
\usepackage{booktabs}
\usepackage{siunitx}
\usepackage{breqn}
\biboptions{square,comma,sort&compress}
\usepackage{hyperref}
\usepackage{float}
\usepackage{subfiles}
\journal{}

\begin{document}

\begin{frontmatter}

\title{Magnetic Field-assisted Crystallographic Selection During Grain Growth in Non-Magnetic Polycrystals: A Phase-Field Study} 

\author[a]{Rakesh Maurya\fnref{equal}}
\author[b]{Soumya Bandyopadhyay\fnref{equal}}
\author[a]{Somnath Bhowmick\corref{cor}*}
\ead{bsomnath@iitk.ac.in}
\author[a]{Rajdip Mukherjee\corref{cor}*}
\ead{rajdipm@iitk.ac.in}
\cortext[1]{Corresponding author}
\fntext[equal]{These authors contributed equally to this work}

\affiliation[a]{Department of Materials Science and Engineering, Indian Institute of Technology, Kanpur, Kanpur 208016, UP, India}
\affiliation[b]{Department of Materials Science and Engineering, University of Florida, Gainesville, Florida 32611, United States}







\begin{abstract}
Magnetic fields drive grain boundary migration in non-magnetic systems through anisotropic susceptibility, yet how the local orientation can affect the resulting collective grain growth behavior remains unclear. We employ a phase-field model to investigate magnetic-field-assisted grain growth in the diamagnetic Bi. This study provides a mechanistic basis for using static and dynamically varied magnetic fields to control texture development and grain growth kinetics in weakly magnetic materials. We also identify the conditions under which magnetic-field-driven orientation selection produces a distinct bimodal grain-size distribution. The bicrystal studies establish that boundary migration is governed by the magnetic free-energy difference between differently oriented grains, while polycrystalline field-sweep simulations reveal a transition from capillary-dominated coarsening to magnetically biased texture selection. For unbiased orientation distributions, magnetic selection remains distributed; however, when a small population of grains is strongly favored, the driving force localizes and exhibits bimodal coarsening characterized by enlarged favored grains embedded within a fine-grained matrix, even without grain boundary energy anisotropy, particle pinning, or heterogeneity. The field removal and field reorientation simulations further show that magnetic field history can redirect the coarsening pathway.

\end{abstract}


\end{frontmatter}

\begin{graphicalabstract}
\includegraphics[width=\textwidth]{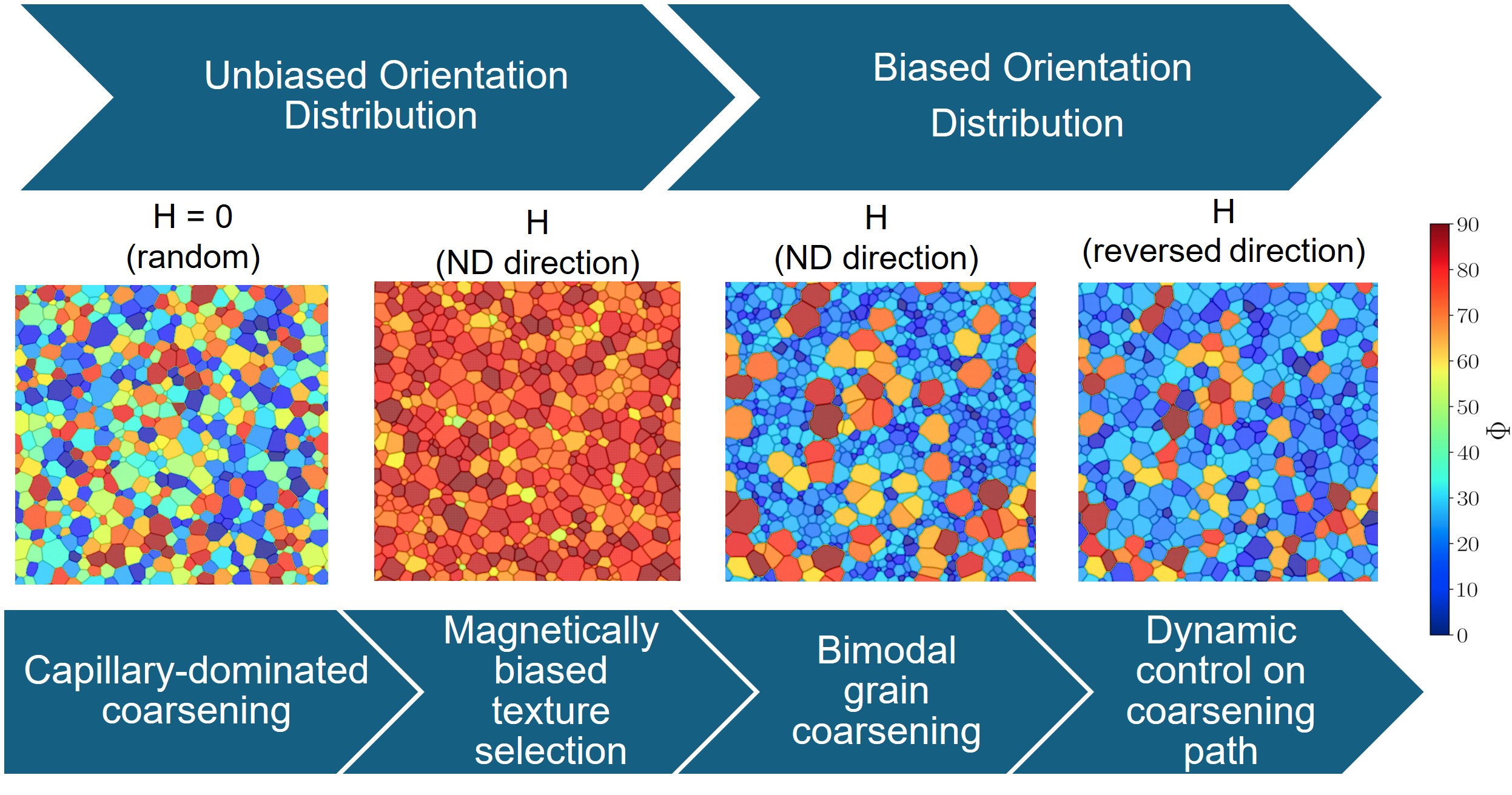}
\end{graphicalabstract}

\section{Introduction}
The continuous increase in the demand for advanced materials with optimized properties has increased the development and deployment of advanced materials processing technologies~\cite{ahmad2012prototyping}.
Recent advances in superconducting magnets have enabled the generation and application of strong magnetic fields, facilitating magnetic field-assisted processing across various domains of materials science and engineering \cite{ludtka2013strategic, ludtka2012prototyping, ahmad2012prototyping}. Studies show that magnetically processed steels demonstrate accelerated transformation kinetics~\cite{ludtka2004situ,ahmad2012prototyping} with improved tensile strength and hardness~\cite{ludtka2012prototyping}. Furthermore, significant improvements in microstructural features, property enhancements in welded joints, and remarkable improvement in the ultimate tensile strength (UTS) are achieved through the application of magnetic field-assisted arc welding~\cite{wu2017review,chen2020effect}. 


Additionally, magnetic fields can influence solidification processes through magnetothermodynamic effects, controlling macroscopic segregation and orientation to reduce defects and inclusions~\cite{yu2020atmospheric,dutta2020effect}. Hou \textit{et al.}~\cite{hou2020control} experimentally studied the impact of axial magnetic fields on the structure of directionally solidified Ni-Mn-Ga alloys, where they obtained a unique composite structure consisting of columnar and equiaxed grains, which is nearly unattainable without an external magnetic field. Furthermore, their report shows improved mechanical performance, indicating that combining magnetic field-assisted solidification with compressive loading can be an effective approach to designing microstructures that optimize material response.

Since Fe or Ni-based alloys are inherently ferromagnetic, they are naturally susceptible to external magnetic fields. However, magnetic field-induced morphological modifications are not only restricted to the ferromagnetic systems. Even, several scientific reports suggest that an external magnetic field can be used efficiently as an additional control over the development of texture and microstructure evolution in various diamagnetic as well as paramagnetic materials~\cite{mullins1956magnetically, molodov2004effect, molodov2005magnetically}. For instance, Cai \textit{et al.}~\cite{cai2019effects} reported that application of a static magnetic field during thermal treatment refines long-period stacking ordered laths that enhance the compressive strength of MgAl-Gd alloys. Another report by Li \textit{et al.} suggests application of rotating magnetic field improves eutectic phase distribution and reduces dendrite arm spacing in solidified Al-7 wt\% Si alloys~\cite{li2020modification}. 

Additionally, the magnetic field can even influence the motion of grain boundaries (GBs). Grain growth (GG) in strain-free grains is driven by grain boundary curvature, which minimizes the overall grain boundary (GB) energy. In magnetically anisotropic materials, the evolution of microstructure is influenced by an additional driving force arising from the varying free energy density of the magnetic susceptibility of the grains \cite{mullins1956magnetically}. Furthermore, the magnetic driving force depends on the applied field and its orientation relative to neighboring grains. In their classic experimental report, Modolov et al. discussed the grain boundary motion during grain growth in non-magnetic Bismuth (Bi)~\cite{molodov1997motion}. Similar behavior was also reported in Zinc (Zn), Titanium (Ti), etc. \cite{molodov2022grain, molodov2007annealing, molodov2010observations, molodov2014grain, sheikh2003migration}. These studies demonstrate that, similar to ferromagnetic materials, magnetic fields can alter microstructural characteristics in various non-ferromagnetic materials and accelerate microstructural development during annealing.


Phase-field modeling has been widely used for simulating phase transformations under external magnetic fields~\cite{zeng2022gibbs, koyama2004phase, lv2010phase, CHAFLE2022132630}. Koyama and Onodera developed a phase-field model of twin microstructure evolution in Ni$_{2}$MnGa subjected to both stress and magnetic fields~\cite{koyama2003phase}. Later, they extended their model to TMP of $\textrm{Fe-Cr-Co}$ alloys~\cite{koyama2004phase}. Sun \textit{et al.}~\cite{sun2013experimental} developed a model of AlNiCo alloys with an applied magnetic field. Zeng \textit{et al.}~\cite{zeng2022gibbs} developed a phase field model of the formation and elongation of $\gamma$ particles in a $\alpha$ matrix in steels. Recently, Bandyopadhyay \textit{et al.}~\cite{bandyopadhyay2026effect} explored the combined chemo-magneto-elastic effects in plain carbon steel. All the studies show the preferential elongation and directionality toward the applied field.

Furthermore, several numerical studies \cite{goins2018effect, rezaei2021phase, liang2022phase} have also explored the influence of magnetic field on microstructural evolution of non-magnetic systems. Rezaei et al. \cite{rezaei2021phase} investigated the effects of magnetic field intensity and direction on the evolution of microstructure and polycrystalline texture in Ti using the phase field method. Their work demonstrated that magnetic fields could significantly alter grain growth kinetics and final texture characteristics, validating theoretical predictions of orientation-dependent growth. Recent studies by Bandyopadhyay~\textit{et al.} \cite{bandyopadhyay2023effect, bandyopadhyay2025grain} have examined magnetic field-driven grain boundary migration in non-magnetic Bi systems. They show that grains with their c-axis aligned parallel to the applied magnetic field exhibit preferential growth due to reduced magnetic free energy. Their model efficiently captured both planar and curvature-driven boundary migration~\cite{bandyopadhyay2023effect}. 

Despite these advances, the current understanding of magnetic field-assisted GG in non-magnetic systems remains largely centered on static-field conditions, in which the magnitude and direction of the applied field are held fixed throughout microstructural evolution. Such studies have been essential for establishing the role of magnetic susceptibility anisotropy in producing orientation-dependent GB migration. However, it remains unclear how this local orientation selection mechanism translates into collective GG behavior in polycrystals, where multiple grain orientations, boundary curvature, junction topology, and initial texture distributions interact simultaneously. In particular, the conditions under which magnetic orientation selection produces only texture-biased normal coarsening, or instead becomes sufficiently localized to trigger distinct bimodal characteristics, remain poorly understood. Furthermore, it is not yet clear whether the growth pathway established under one field direction can be modified by removing or reorienting the magnetic field during microstructural evolution.

Addressing this question is important for both fundamental and applied reasons. From a scientific perspective, this clarifies how magnetic susceptibility anisotropy interacts with initial texture and orientation bias to produce selective grain growth, and whether this coupling can serve as an independent thermodynamic route to achieve bimodal characteristics without assistance from mechanisms such as particle pinning, GB energy anisotropy, or mobility anisotropy. From a processing perspective, it provides guidance for using magnetic fields not only to accelerate grain growth but also to control texture development and avoid unwanted microstructural instability.

Motivated by this, we develop a comprehensive phase-field model that incorporates magnetic field effects to study GB migration, using diamagnetic polycrystalline Bi as a prototype system. The computational model accounts for crystallographic orientation-dependent magnetic free energy, enabling us to examine how field magnitude, field direction, initial texture, and field history influence magnetic orientation selection and the resulting microstructural evolution.

\section{Model formulation}\label{sec:model}
\subsection{Phase-field model}
Over the past few decades, the phase-field method has established itself as a robust computational framework for predicting mesoscale microstructural evolution based on thermodynamics and kinetics of materials \cite{moelans2008introduction}. A key advantage of this approach is its ability to capture microstructure evolution under external stimuli without imposing any a priori assumptions on the system morphology \cite{boettinger2002phase,chen2002phase,guin2025interplay, bandyopadhyay2023role, bandyopadhyay2023multivariant, chafle2019effect}. Diffuse interfaces between grains are described implicitly via spatial gradients of these order parameters, and their evolution, driven by thermodynamic forces, captures microstructural changes via the motion of diffuse grain boundaries.  In this study, we adopt a non-conserved order parameter formalism, where a set of order parameter fields $\eta_i(\vec{r})$, i = 1, 2, \ldots, N, each associated with a grain of distinct crystallographic orientation, evolves over the simulation domain \cite{moelans2008introduction}.

The total free energy functional ($F$) of polycrystalline material is defined as :~\cite{cahn1958free, jamshidian2014phase}.
\begin{equation}
F = \int_V \left( u f_{(0)}(\eta_i) + \kappa_\eta \sum_{i=1}^N |\nabla \eta_i|^2 + f_m \right) dV,
\label{eq:1}
\end{equation}
where the parameters $u$ and $\kappa_{\eta}$ are the bulk free energy density and gradient energy coefficients, respectively, and $f_{m}$ is the magnetic free energy density. These coefficients are related to physically measurable GB properties; the specific GB energy $ \sigma_{GB} = 0.155 J/m^2$~\cite{ocak2008interfacial} and the diffuse interface interfacial width $l_{GB}$ through the relations $l_{GB}$ as  $u = 6\sigma_{GB}/l_{GB}$ and $\kappa_{\eta}= 3\sigma_{GB}l_{GB}/4$ \cite{bandyopadhyay2023effect, bandyopadhyay2025grain}. All three terms in Eq.~\eqref{eq:1} represent, respectively, the free energy density, the gradient (interfacial) energy penalizing spatial variation in the order parameters, and the orientation-dependent magnetic free energy.

The bulk free energy density for a polycrystalline material with N distinct grain orientations is given as \cite{rezaei2021phase, liang2022phase, moelans2008introduction}: 
\begin{equation}
f_{(0)}(\eta_i) = \sum_{i=1}^N \left( \frac{\eta_i^4}{4} - \frac{\eta_i^2}{2} \right) + \gamma \sum_{i=1}^N \sum_{j>i} \eta_i^2 \eta_j^2 + 0.25,
\label{eq:2}
\end{equation}
The first term of Eq.~\eqref{eq:2} contains the standard double well potential in each order parameter, with minima at $\eta_{i} = 0$ and $\eta_{i} = 1$ corresponding to the absence and presence of grain $i$, respectively. The second term, weighted by the interaction parameter $\gamma=1.5$, penalizes the coexistence of multiple order parameters at the same spatial location, thereby confining each grain boundary to a finite diffuse thickness and preventing spurious multi-phase junctions. The constant 0.25 ensures that the energy minimum inside a grain is zero \cite{rezaei2021phase, moelans2008introduction}.

Additionally, we formulate the magnetic free energy density, which couples grain crystallography to the external field, by weighting per grain magnetic energy density through a smooth interpolation (switching) function
\cite{allen2016simulations}:
\begin{equation}
f_m = \sum_{i=1}^N h(\eta_i) f_{m,i}~,
\label{eq:3}
\end{equation}
where the switching function $h(\eta_i) = \eta_i^3 (6.0 \eta_i^2 - 15.0\eta_i +10)$ varies from 0 to 1 as $\eta_i$ transitions from 0 to 1. This ensures that the magnetic energy contribution is only active within the bulk of each grain. Thus the magnetic free energy density $f_{m,i}$ is given by \cite{rezaei2021phase,molodov2014grain}:
\begin{equation}
f_{m,i} = \frac{1}{2}\mu_{0}H^{2}\chi_{i}~,
\label{eq:4}
\end{equation}
where $H$ is the magnitude of the applied magnetic field vector, $\mu_{0} = 1.257 \times 10^{-6}\ \mathrm{N\,A^{-2}}$ is the permeability of free space, and $\chi_{i} $ is the effective magnetic susceptibility of grain \(i\) projected along the field direction. This scalar quantity depends on both the intrinsic susceptibility tensor of the crystal and the crystallographic orientation of the grain relative to the applied field.

\begin{figure}[htbp]
    \centering    \includegraphics[width=0.6\linewidth]{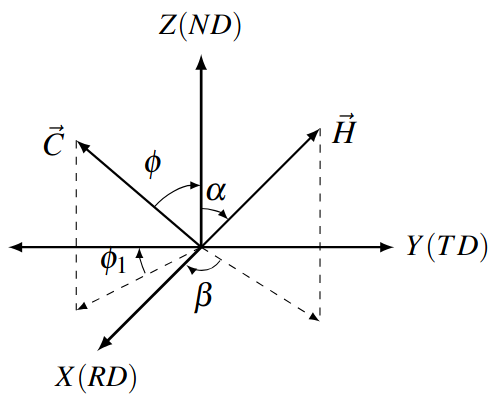}  
    \caption{Orientation of the magnetic field and principal axes with respect to rolling direction (RD), transverse direction (TD), and Normal direction (ND). }
     \label{fig:plane}
\end{figure}

In this study, we consider a diamagnetic metal, Bi, with a rhombohedral crystal structure that can be approximated as hexagonal for magnetic susceptibility analysis~\cite{molodov1997motion}. Due to hexagonal symmetry consideration, the magnetic susceptibility tensor reduces to two independent components: $\chi_{a}=\chi_{b}$ in the basal plane ($a_{1}, a_{2}, a_{3}$) directions and $\chi_{c}$ along the principal c-axis. The effective susceptibility of grain $i$ under an applied field is defined as \cite{asai2003crystal,molodov2014grain}:
\begin{equation}
\chi_{i} = \chi_{a} + (\chi_{c}-\chi_{a}) \cos^{2}(\gamma_{i}),
\label{eq:5}
\end{equation}
where $\gamma_{i}$ is the angle between the applied magnetic field vector ($\vec{H}$) and the principal c-axis ($\vec{C_{i}}$) of grain $i$. The cosine of this angle is computed from the dot product as given in \cite{allen2016simulations}:
\begin{equation}
\cos(\gamma_i) = \frac{\vec{C} \cdot \vec{H}}{|\vec{C}| \, |\vec{H}|}.
\label{eq:6}
\end{equation}
Thus, the direction of the c-axis for each grain in the sample reference frame is obtained by applying the Bunge rotation matrix, parameterized by the Euler angles ($\phi_{1},\Phi, \phi_{2}$), to the c-axis vector of the crystal coordinate system. We define the three successive rotations that transform the crystal frame into the sample frame by the rolling direction (RD), transverse direction (TD), and normal direction (ND) (see Fig.~\ref{fig:plane}.). In component form, we express the c-axis vector in the sample frame as:
\begin{equation}
\vec{C} = \big( C \sin\Phi \sin\phi_{1},\; -C \sin\Phi \cos\phi_{1},\; C \cos\Phi \big).
\label{eq:7}
\end{equation}
Therefore, we characterize the applied magnetic field direction by two angles: $\alpha$, the angle between $\vec{H}$ and the ND, and $\beta$, the angle between the RD and the projection of $\vec{H}$ onto the RD--TD plane. In all polycrystalline simulations shown in this work, we constrain the field to lie in the ND--TD plane, corresponding to $\beta = 90^\circ$. Thus, the field vector is given as:
\begin{equation}
\vec{H} = ( H \sin\alpha \cos\beta, \; H \sin\alpha \sin\beta, \; H \cos\alpha ).
\label{eq:8}
\end{equation}
Substituting Eqs.~\eqref{eq:7} and~\eqref{eq:8} into Eq.~\eqref{eq:6}, the cosine of the angle between the field and the c-axis of grain i, is simplified as:
\begin{equation}
\cos(\gamma_{i}) =\sin\alpha \cos \beta \sin \Phi \sin \phi_{1}
 - \sin \alpha \sin \beta \sin \Phi \cos \phi_{1}
 + \cos \alpha \cos \Phi ,
 \label{eq:9}
\end{equation}
This expression establishes a direct link between the macroscopic field direction ($\alpha,\beta$) and the microscopic grain orientation ($\phi_{1},\Phi, \phi_{2}$), enabling the computation of the orientation-specific magnetic driving force for each grain in the polycrystalline materials.

The spatiotemporal evolution of each non-conserved order parameter is governed by the Allen-Cahn (AC) equation, obtained by minimizing the total free energy functional with respect to $\eta_i$ \cite{krill2002computer,chen2002phase}.
\begin{equation}
\frac{\partial \eta_i}{\partial t} = -L \frac{\delta F}{\delta \eta_i} 
 \label{eq:10}
\end{equation}
Substituting Eq.~\eqref{eq:1} into the variational derivative, the strong form evolution equation for order parameter $\eta_{i}$ becomes:
\begin{equation}
\frac{\partial \eta_i}{\partial t} = -L \left[
u \left( \eta_i^3 - \eta_i + 2 \gamma \eta_i \sum_{j \ne i}^{N} \eta_j^2 \right)
- \kappa_\eta \nabla^2 \eta_i 
+ \frac{\partial f_{m,i}}{\partial \eta_i}
\right]
 \label{eq:11}
\end{equation}
where the relaxation parameter $L = \frac{4m}{3l_{gb}}$, and $m = m_0 \exp(-Q/k_BT)$, with $m_0$ being the pre-exponential factor, $Q$ the activation energy, and $k_B$ the Boltzmann constant.

\section{Simulation Details} \label{sec:sim_details}
\subsection{Model Implementation in MOOSE} 
We solve the system of partial differential equations mentioned in the previous section using the finite element method (FEM) and implicit time integration with the Multiphysics Object-Oriented Simulation Environment (MOOSE), an open source FEM framework \cite{permann2020moose}. We evolve the grain order parameters (Eq.~\eqref{eq:11}), using  Newton's method~\cite{permann2020moose} with a second-order backward Euler differentiation scheme. The linear subproblems are handled by the GMRES solver, preconditioned with block Jacobi. The spatial domain is discretized using a structured quadrilateral mesh of QUAD4 elements. We employ adaptive time-stepping with an initial time-step size of 0.01 and roughly 15 nonlinear iterations per time step.  Additionally, we employ automatic nonlinear residual scaling and set the relative and absolute convergence criteria to $10^{-8}$, ensuring numerical robustness throughout each simulation.

\subsection{Simulation setup} 
We choose the interfacial width of $l = 5\delta$, where $\delta = 0.32$ nm is the smallest grid spacing. We employ dynamic mesh adaptivity to concentrate resolution where it matters most, thereby reducing computational cost.  At each time step, a local error is computed for every mesh element, flagging regions with large gradients (typically around interfaces) for refinement and regions with no gradients for coarsening.
\begin{figure}[H]
    \centering    \includegraphics[width=1.0\linewidth]{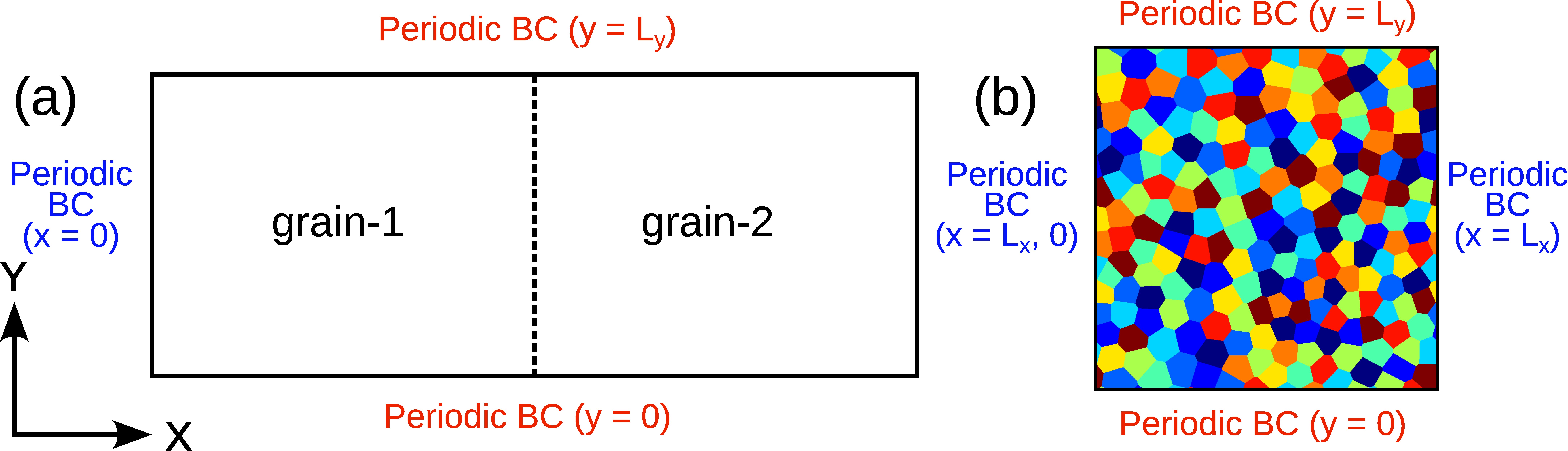}  
   \caption{Schematic showing the boundary conditions (BCs). In both the bicrystal system [Fig.~\ref{fig:bc}(a)] and the polycrystalline system [Fig.~\ref{fig:bc}(b)], periodic boundary [ $\eta_{i}(x=0,y)=\eta_{i}(x=L_x,y)$ and $\eta_{i}(x,y = 0)=\eta_{i}(x,y=L_y)$] conditions are applied in the x- and y- directions respectively. Here, $\eta_{i}$ denotes a set of non-conserved order parameters.}
     \label{fig:bc}
\end{figure}

For the bicrystal simulations, we impose periodic boundary conditions along both directions to eliminate artificial edge effects on the GB. The computational domain is set to $1024\times 128$, with two grains initialized such that a flat GB lies at the center of the domain as shown in Fig.~\ref{fig:bc}(a). This geometry is the standard configuration for studying GB migration under an applied magnetic field, and the flat interface ensures that the capillary contribution remains minimal, providing a clean baseline for measuring magnetically driven migration. 

For the polycrystalline simulations in Fig.~\ref{fig:bc}(b), we consider a periodic computational domain of $2048 \times 2048$ to accommodate a statistically representative initial population of 4000 grains generated by Voronoi tessellation. The grains are uniformly distributed at the initial stage to avoid spatial or orientational bias in the starting microstructure.

The magnetic material parameters used in Eq.~\eqref{eq:4} are assigned values consistent with those reported in the experimental literature for Bi~\cite{asai2003crystal,bandyopadhyay2025grain,molodov2010observations,liang2022phase}. The permeability of free space is $\mu_0 = 1.257 \times 10^{-6}\ \mathrm{N\,A^{-2}}$, and the magnetic susceptibility values along the principal crystallographic axes are taken as $\chi_a = \chi_b = -1.24 \times 10^{-4}$ and $\chi_c =-1.47 \times 10^{-4}$, reflecting the anisotropic diamagnetic response of Bi. We non-dimensionalize all the material parameters following Ref.~\cite{bandyopadhyay2025grain} as detailed in~\ref{appendix_dimensionless}, and the dimensionless material parameters used in the simulations are listed in Table~\ref{tab:table2}.



\begin{table}[tbph]
\centering
\caption{List of parameters used in this study}
\label{tab:table1}
\begin{tabular}{|l|l|l|}
\hline
\label{tab:table2}
Type & Parameter name & Value \\
\hline
Materials Parameters
& Bulk free energy parameters ($u$, $\gamma$) & 0.8131, 1.5 \\
& Gradient energy coefficient ($\kappa_{\eta}$) & 5.535 \\
& GB relaxation coefficient ($L$) & 1.0\\
\hline
Simulation parameters 
& Simulation box size & $1024 \times 128$ (Bicrystal)\\
& & $2048 \times 2048$ (Polycrystal)\\
\hline
\end{tabular}
\end{table}

\section{Results} \label{sec:results}
We first apply the phase-field model to a 2D bicrystal configuration to establish the fundamental relationship between
grain crystallographic orientation and magnetically driven GB migration. Later, we extend this concept to study the effect of external magnetic fields on non-magnetic crystallographically oriented polycrystals. Note that we perform all simulations assuming isothermal conditions at 498 K.

\subsection{Bicrystalline morphology}\label{bicrystal}
\begin{figure}[h]
    \centering
    \includegraphics[width=1.\linewidth]{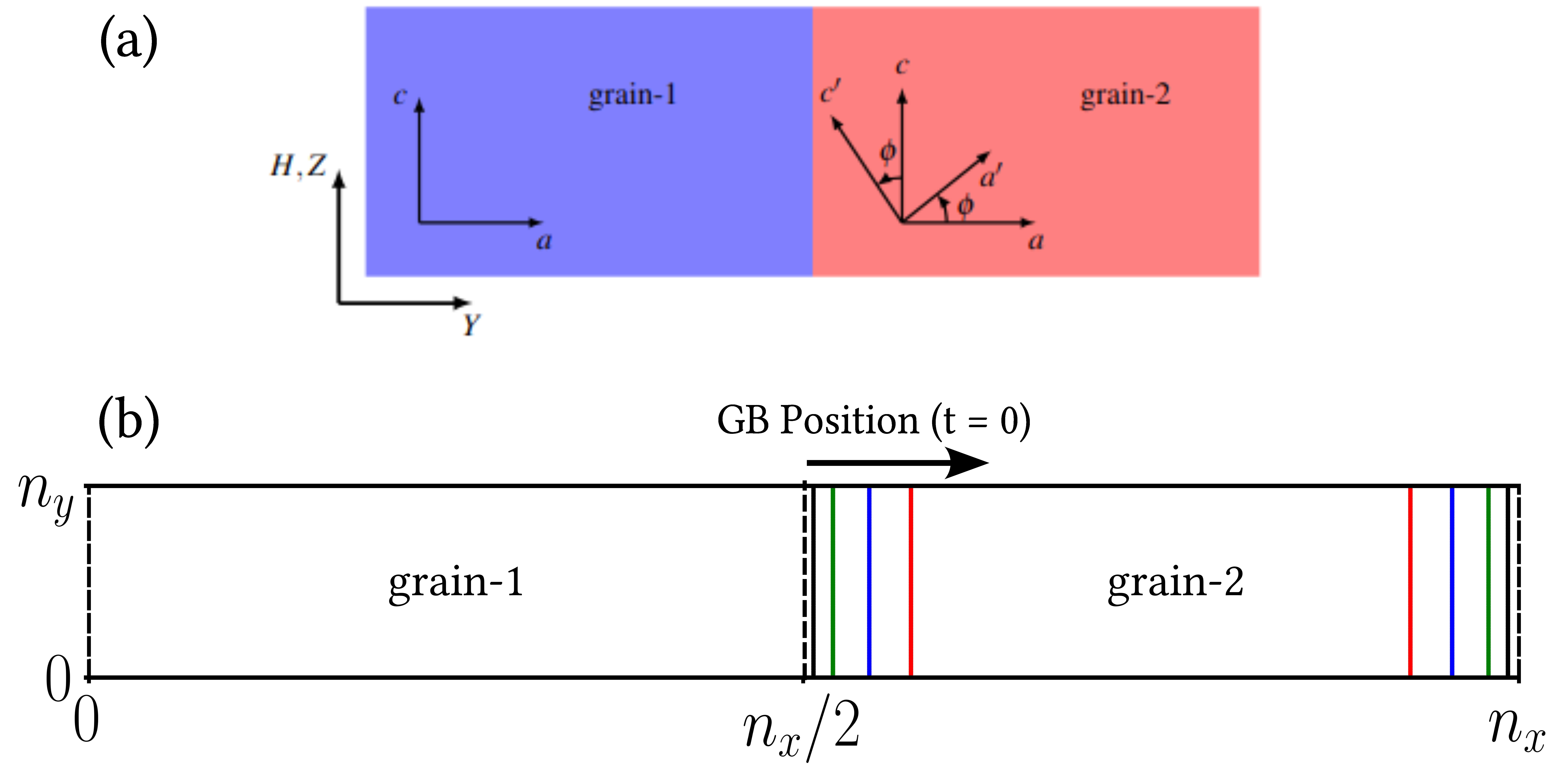} 
    \caption{(a) Schematic representation of a bicrystal system with magnetic field $H_{ext}$ applied along the Z-axis. Grain 1(blue)  is kept fixed, while the orientation of Grain 2 (red) is systematically varied. (b) Dash line shows GB position at t = 0 and the contour plots at $t = 25000$ simulation time corresponds to  $\Phi = 10\degree $(black), $\Phi = 30\degree $(green), $\Phi = 50\degree $(blue) and $\Phi = 90\degree $(red), respectively. The arrow indicates the direction of GB migration.}  
    \label{fig:contour}
\end{figure}

Figure~\ref{fig:contour}(a) represents the schematic of the representative bicrystalline microstructure. We assign Grain~1 (blue) to have a fixed orientation of $\Phi =0^\circ$, with its principal $c$-axis aligned with the applied magnetic field direction along the Cartesian $Z$-axis. On the other hand, we systematically vary the orientation of Grain~2 (red) by rotating its $c$-axis counter-clockwise, as illustrated in Fig.~\ref{fig:contour}(a). Additionally, we apply a magnetic field of $H_{ext} = 1.68 \times 10^8$~A/m, along the Cartesian $Z$-axis, and the GB is initialized as a flat interface at the center of the domain.

When the $c$-axes of both grains are parallel ($\Phi = 0^\circ$), we do not observe any GB migration. However, as we rotate the $c$-axis of Grain~2, the grain boundary begins to migrate, with Grain~1 growing at the expense of Grain~2. To observe the movement of the Grain-1 interface, Fig.~\ref{fig:contour}(b) shows the contour plots at a fixed simulation time of $25000$ for different Euler rotation angles of Grain-2, namely  $\Phi = 10\degree $(black line), $\Phi = 30\degree $(green line), $\Phi = 50\degree $(blue line) and $\Phi = 90\degree $(red line), respectively.

\begin{figure}[tpbh]
    \centering
    \includegraphics[width=1\linewidth]{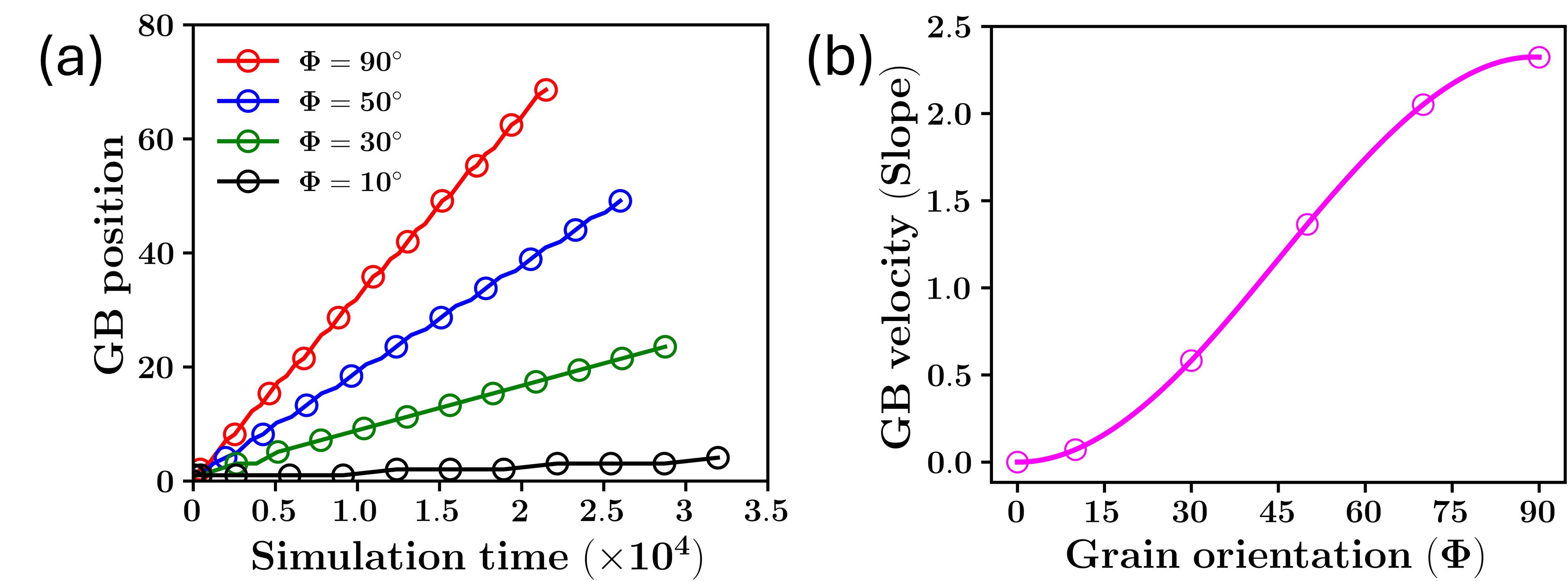}
   \caption{
(a) Temporal evolution of the GB position for different misorientation angles between the crystallographic $c$-axes of both grains. 
(b) GB velocity of Grain~1 as a function of grain orientation. 
}
\label{fig:interface_displacement}
\end{figure}

Furthermore, we plot the corresponding displacement of the GB as a function of time for four representative misorientation angles ($\Phi = 10\degree, 30\degree, 50\degree,$ and $ 90\degree$) in Fig.~\ref{fig:interface_displacement}(a). At $\Phi = 10\degree$, the boundary displacement is minimal (shown by the black curve). As the misorientation increases to $30\degree - 90\degree$, the migration becomes progressively more pronounced as shown by increasing slopes in the time-displacement curves, with the largest boundary displacement observed at $\Phi = 90^\circ$ (red). The corresponding growth rate of Grain~1 (obtained from the slope of the time-displacement curves in Fig.~\ref{fig:interface_displacement}(a)) is plotted as a function of $\Phi$ in Fig.~\ref{fig:interface_displacement}(b). The growth rate increases monotonically with the orientation angle and reaches its maximum at $\phi = 90\degree$, exhibiting a sigmoidal trend. This indicates a gradual transition from weak magnetic bias at small $\Phi$ to strong magnetic-driven migration at larger $\Phi$.

\subsection{Polycrystalline Morphology}\label{polycrystal}
Having established the orientation-dependent magnetic effects on the resulting GB migration behavior in the bicrystal system, we next extend our study to polycrystalline microstructures, where multiple grain orientations and boundary interactions govern the overall evolution.

\subsubsection*{Case 1: Effect of Magnetic Field on Grain Growth}
\begin{figure}[tbph]
\centering
\includegraphics[width=1.05\linewidth]{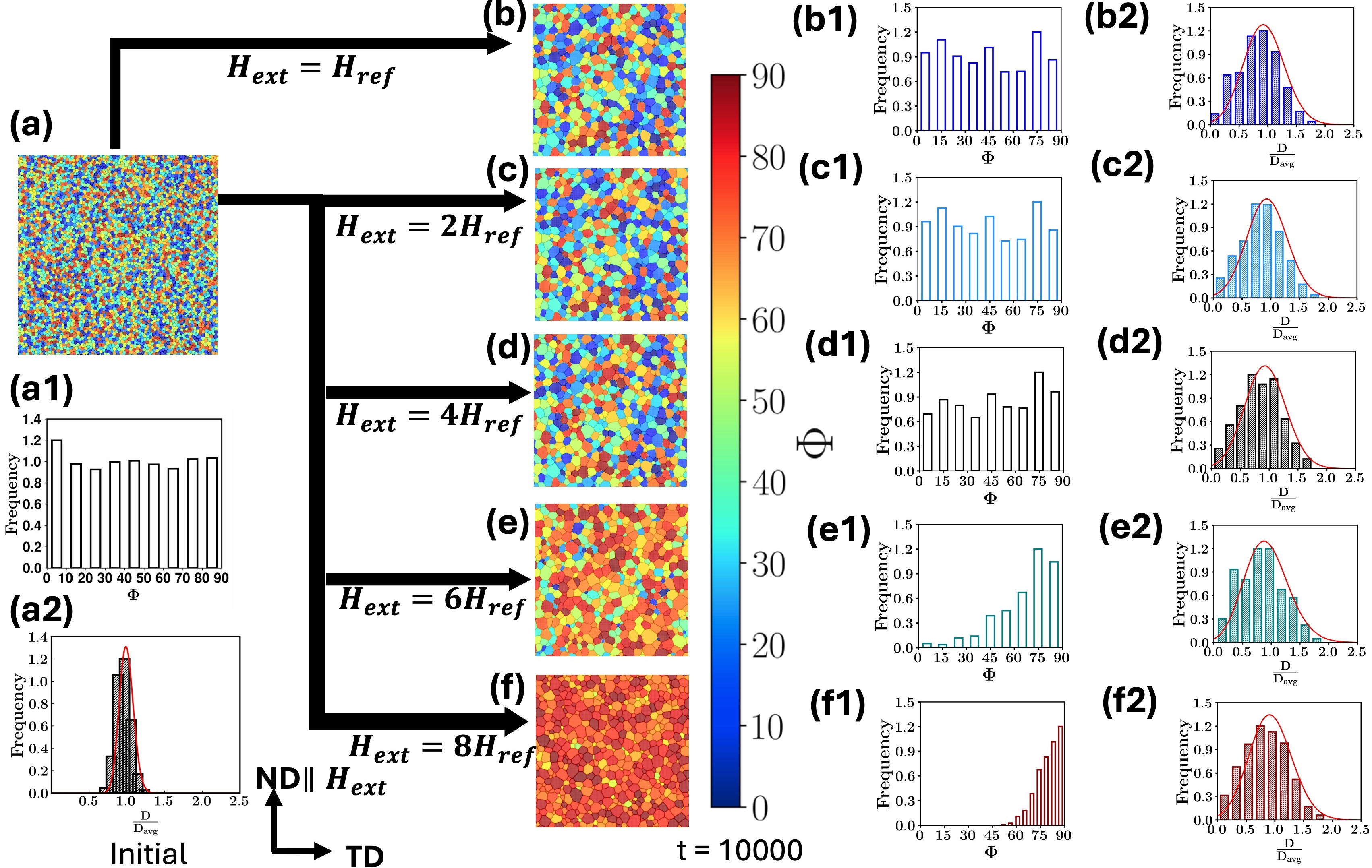} 
\caption{Microstructural evolution, crystallographic orientation distribution, and normalized grain size distribution under magnetic fields applied along the ND direction at different field intensities. (a) Initial microstructure, with the corresponding (a1) Euler angle distribution and (a2) normalized grain-size distribution, respectively. Columns (b)--(f) display the evolved microstructures at simulation time $t = 10000$ for applied magnetic fields of $H_{\mathrm{ref}}$, $2H_{\mathrm{ref}}$, $4H_{\mathrm{ref}}$, $6H_{\mathrm{ref}}$, and $8H_{\mathrm{ref}}$, respectively. The corresponding Euler angle distributions and normalized grain size distributions are shown in columns (b1)--(f1) and (b2)--(f2). The color bar represents the crystallographic orientation of the grains.
}
\label{fig-micropoly}
\end{figure}

First, we investigate the effect of the magnetic field on the polycrystal morphology, which is similar to that shown in Sec .~\ref {bicrystal} for the bicrystal system. We apply the external magnetic field along the ND ($\alpha = 0^\circ$ in Fig.~\ref{fig:plane}) direction to the ND--TD plane ($\beta = 90^\circ$) of the sample. We initialize the configuration with 4000 grains generated by Voronoi tessellation as shown in Fig.~\ref{fig-micropoly}(a), with Euler angles assigned randomly and uniformly over the range $0^\circ$--$90^\circ$ (See Fig.~\ref{fig-micropoly}(a1)). Additionally, the normalized grain size distribution (expressed in terms of $D_{\max}/D_{\text{avg}}$) is relatively narrow and centered around $\approx 1$. This indicates the unbiased nature of the initial microstructure. Furthermore, the external field of $1.68\times 10^8$ A/m, used to study GB migration for the bicrystal system (Sec.~\ref{bicrystal}) is considered as the reference magnetic field $H_{ref}$.

Unlike the bicrystal configuration, where the boundary is essentially planar and curvature effects are minimal, the polycrystalline system contains a network of curved GBs and different junctions. As a result, the microstructural evolution is governed by the coupled interplay between the capillarity-driven GG and magnetic contribution. When the capillarity contribution is stronger than its magnetic counterpart, the microstructure evolution will mainly be curvature dominated. In contrast, when the magnetic contribution becomes comparable to or larger than the capillary contribution, the field can significantly modify the GG kinetics and the resulting microstructure.

To examine this competition systematically, the applied magnetic field is scaled by the reference field as $H_{\mathrm{ext}} = nH_{\mathrm{ref}}$, where $n$ is varied over the selected range from 1 to 9. This allows us to identify the field strength regime in which the magnetic contribution begins to noticeably alter the polycrystalline GG behavior relative to capillarity-dominated coarsening. The field magnitude study displays a threshold-like response rather than a gradual linear change in morphology. For $H_{\mathrm{ext}} = H_{\mathrm{ref}}, 2H_{\mathrm{ref}}, \;\mathrm{and} \; H_{\mathrm{ext}} = 4H_{\mathrm{ref}}$, we do not observe any significant difference in the morphology even after 10000 simulation time steps as shown in Figs.~\ref{fig-micropoly}(b), (c), and (d), respectively. Furthermore, the corresponding orientation distributions do not show a strong deviation from the initially uniform distribution (see Figs.~\ref{fig-micropoly}(b1), (d1), (e1)). The normalized grain size distributions also remain centered near $D_{\max}/D_{\text{avg}} \approx 1$.

Nevertheless, a clear deviation appears when the field magnitude is increased to $6H_{\mathrm{ref}}$. The microstructure also begins to show a more pronounced enrichment of grains with magnetically favored orientations (See Fig.~\ref{fig-micropoly}(e)). At this field strength, the orientation distribution becomes strongly biased toward higher $\Phi$ as shown in Fig.~\ref{fig-micropoly}(e1). This indicates that magnetic selection has become significant. This also suggests that the magnetic driving force has reached a level comparable to the capillary contribution and can now noticeably influence the collective GG behavior. At $8H_{\mathrm{ref}}$, the magnetic effect becomes significantly dominant. The microstructure is largely enriched in high $\Phi$ grains, and the orientation distribution is strongly concentrated near $70^\circ–90^\circ$ (See Figs.~\ref{fig-micropoly}(f) and (f1)). 

However, the normalized grain size distribution remains broadly comparable across the field magnitudes. Although we observe a minor broadening at higher fields, the distributions remain centered around $D/D_{\mathrm{avg}}\approx 1$ and do not develop a pronounced long tail toward large grain sizes (shown in Figs.~\ref{fig-micropoly}(e2),(f2)).
This indicates that increasing the field magnitude in an initially unbiased orientation distribution primarily promotes the selection of crystallographic texture. 

\begin{figure}
    \centering
    \includegraphics[width=1.0\linewidth]{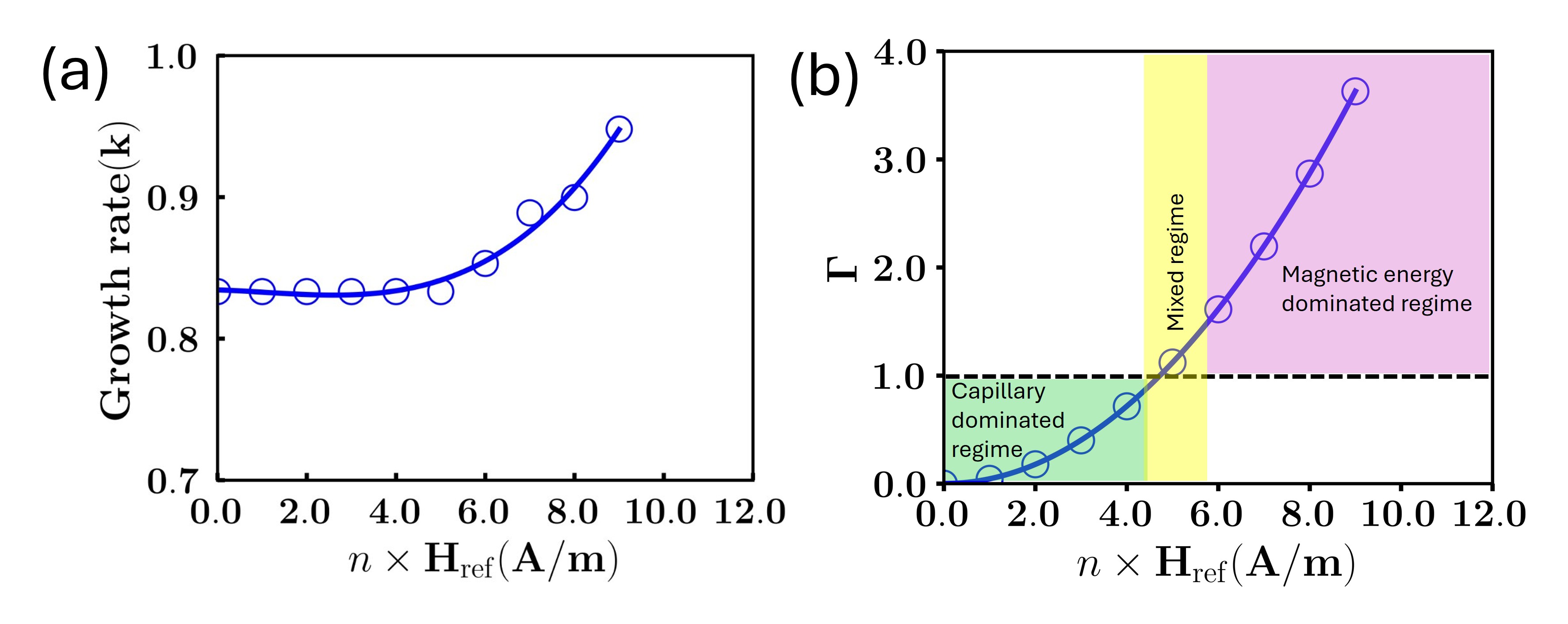} 
 \caption{(a) Growth rate, quantified from the slope of the average squared grain size $\langle D^2 \rangle$ versus simulation time, for different applied magnetic fields. (b) Variation of the magnetic to capillary driving force ratio, $\Gamma$, with applied magnetic field strength. The horizontal dashed line at $\Gamma = 1$ denotes comparable magnetic and capillary driving forces. The shaded regions indicate the capillarity-dominated, mixed, and magnetic-energy-dominated regimes, respectively, that govern microstructural evolution.
} \label{magnetic_to_capillary_ration1}
\end{figure}

These responses are further confirmed by the slope or growth rate constant (k) plot of $\langle D^2\rangle$ vs simulation time as a function of magnetic field, displayed in Figure~\ref{magnetic_to_capillary_ration1}(a). For $H_{\mathrm{ext}}\leq4.7H_{\mathrm{ref}}$, k remains nearly unchanged, indicating that the coarsening kinetics are still primarily capillarity dominated. A clear increase in k is observed only beyond $6H_{\mathrm{ref}}$, consistent with the onset of appreciable magnetic contribution to the grain growth kinetics. This transition coincides with the emergence of a strong shift in the orientation distribution toward high $\Phi$ grains discussed above. To further confirm this, we plot the magnetic energy to capillary or curvature induced interfacial energy ratio ($\Gamma$) as a function of magnetic field, as shown in Fig.~\ref{magnetic_to_capillary_ration1}(b). When $\Gamma \ll 1$ or small, i.e., when the applied external magnetic field $H_{ext} < 4.7H_{ref}$, the evolution remains predominantly curvature driven; when it becomes sufficiently large ($\Gamma \gg 1$) and the magnetic field is $H_{ext} > 6H_{ref}$, the magnetic field can modify the growth kinetics and bias the evolving texture. When $\Gamma \approx 1$ the system is mixed controlled at $4.7H_{ref} \lesssim H_{ext} < 6H_{ref}$. This trend is consistent with our polycrystalline simulations displayed in Fig.~\ref{fig-micropoly}. We discuss the detailed calculation of $\Gamma$ in ~\ref{appn}.

After identifying $8H_{\mathrm{ref}}$ as a field magnitude that produces strong magnetic texture selection, we next examine the role of field direction by applying the same field magnitude along two orthogonal directions, ND and TD. This comparison isolates the directional nature of the magnetic driving force while keeping the field strength fixed.

\begin{figure}
\centering
\includegraphics[width=0.9\linewidth]{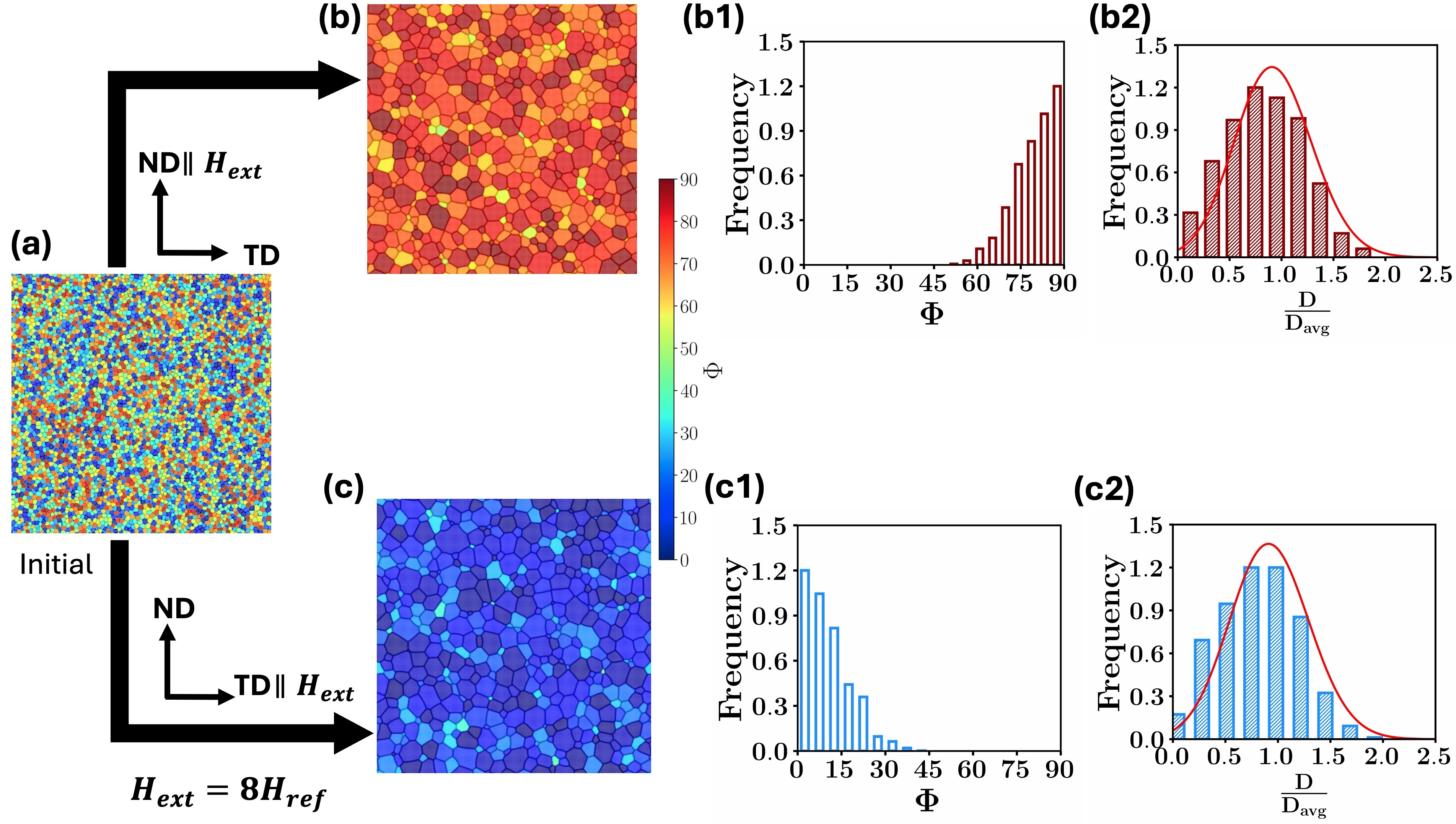} 
\caption{Microstructural evolution under an applied magnetic field of $H_{ext} = 8H_{\mathrm{ref}}$ along different field directions: (a) Initial microstructure at simulation time $t=0$, (b) and (c) evolved microstructures at $t = 10000$ for magnetic fields applied along the ND and TD directions, respectively. The corresponding Euler angle and normalized grain size distributions are displayed in columns (b1)--(c1) and (b2)--(c2). The color bar represents the crystallographic orientation of the grains. For both cases, the continuous temporal evolution of the microstructure, along with the corresponding evolution of the Euler angles and grain size distribution, can be observed in the supplementary video files.}
\label{fig-microstructure8href}
\end{figure}

When $H_{\mathrm{ext}}\parallel$ ND, the evolved microstructure becomes strongly favored by high $\Phi$ grains, as shown in Fig.~\ref{fig-microstructure8href}(b). The corresponding orientation distribution in Fig.~\ref{fig-microstructure8href}(b1) shifts toward $\Phi \approx 70^\circ–90^\circ$, indicating preferential growth of grains whose crystallographic orientation is favored under the field along ND. On the other hand, when we apply the same field along TD, we observed an opposite trend. The evolved microstructure in Fig.~\ref{fig-microstructure8href}(c) is dominated by low $\Phi$ grains, and the orientation distribution in Fig.~\ref{fig-microstructure8href}(c1) shifts toward $\Phi \approx 0^\circ–20^\circ$. This also confirms that reversing the field direction changes the magnetic free-energy hierarchy among grain orientations, thereby selecting a different population of orientations.

\begin{figure}[ht]
    \centering
    \includegraphics[width=0.6\linewidth]{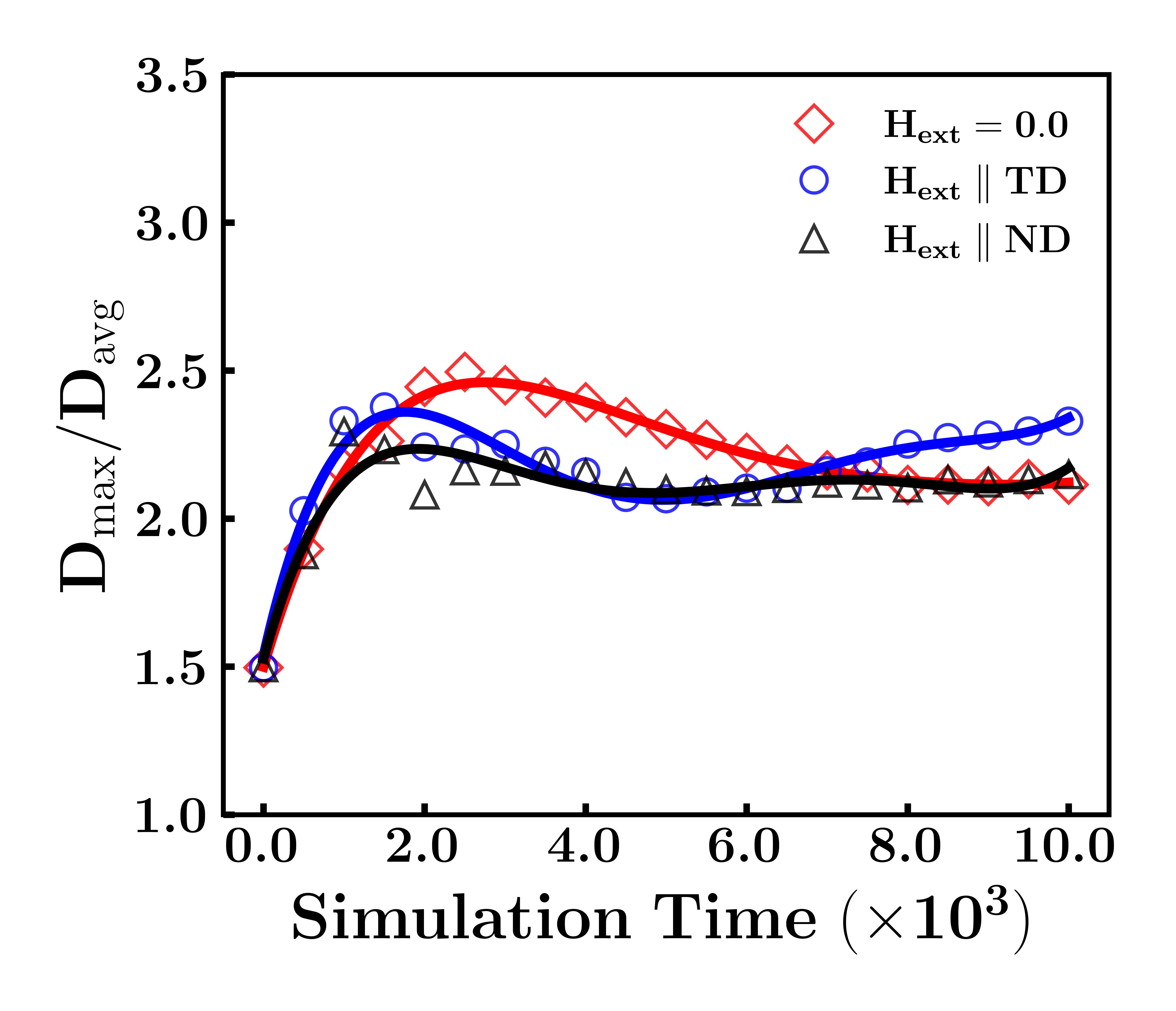} 
    \caption{Evolution of the normalized maximum grain size ratio, $D_{\max}/D_{\mathrm{avg}}$, as a function of simulation time for no field (red), $H_{\mathrm{ext}} = 8H_{\mathrm{ref}}$ applied along ND (black), and TD (blue) directions, respectively. The comparable trends indicate that field direction primarily controls the selection of crystallographic texture rather than producing any grain-size anomaly in the initially unbiased polycrystal.}
\label{figmax_D_vs_D_nt_td_without_field}
\end{figure}

Importantly, the normalized grain size distributions in Figs.~\ref{fig-microstructure8href}(b2) and (c2) remain comparable. This indicates that changing the field direction does not produce a drastic change in grain-size statistics; however, it reverses the crystallographic texture during growth. Furthermore, the temporal evolution of $D_{\max}/D_{\mathrm{avg}}$ further supports this interpretation (see Fig.~\ref{figmax_D_vs_D_nt_td_without_field}). Here, $D_{max}$ corresponds to the maximum gain diameter, while $D_{\mathrm{avg}}$ denotes the average grain diameter of the system. Although the ND and TD directed fields select opposite orientations of the grains, their $D_{\max}/D_{\mathrm{avg}}$ values remain broadly comparable to the field-free case. This further indicates that in the initially unbiased polycrystal, the applied magnetic field primarily controls the crystallographic texture rather than producing any anomaly in grain size distribution.

\subsubsection*{Case 2: Role of crystallographic texture and orientation bias}
The field sweep study establishes that, for an initially unbiased polycrystal, increasing the field magnitude primarily produces texture selection and enhanced coarsening, while the grain size distribution remains broadly distributed. This suggests that the magnetic driving force is shared among many favorably oriented grains when such orientations are initially distributed throughout the microstructure. Therefore, to determine whether magnetic selection can become localized onto a limited grain population, we next introduce controlled initial texture biases.

In this section, we further extend our study to understand the role of initial texture and orientation bias on the final morphology subjected to a directionally applied magnetic field. In particular, we investigate whether a small population of magnetically favored grains can grow selectively within a matrix of less favorably oriented grains, and whether this selective growth can trigger bimodal growth. Note that we use $H_{ext}=8H_{ref}$ as an external magnetic field for all the studies discussed below.

To this end, we assign four different sets of orientations as follows:
\begin{itemize}
    \item Set 1: 90\% ($0^\circ$-$45^\circ$) \& 10\% ($45^\circ$-$90^\circ$)
    \item Set 2: 98\% ($0^\circ$-$45^\circ$) \& 2\% ($45^\circ$-$90^\circ$)
    \item Set 3: 98\% ($0^\circ$-$30^\circ$) \& 2\% ($60^\circ$-$90^\circ$)
    \item Set 4: 98\% ($0^\circ$-$10^\circ$) \& 2\% ($80^\circ$-$90^\circ$)
\end{itemize}

\begin{figure*}
   \centering
    \begin{subfigure}[t]{\textwidth}
        \centering
        \includegraphics[width=0.75\linewidth]{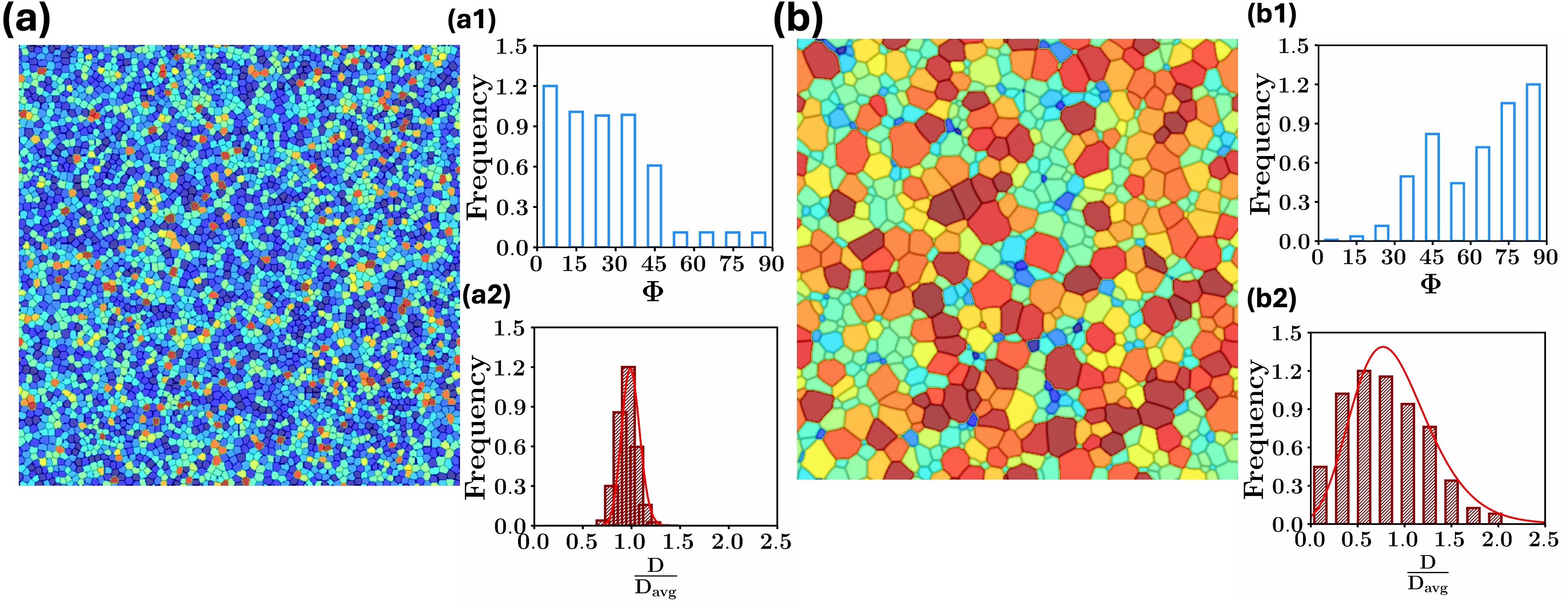}
    \end{subfigure}
    
     \begin{subfigure}[t]{\textwidth}
        \centering
        \includegraphics[width=0.75\linewidth]{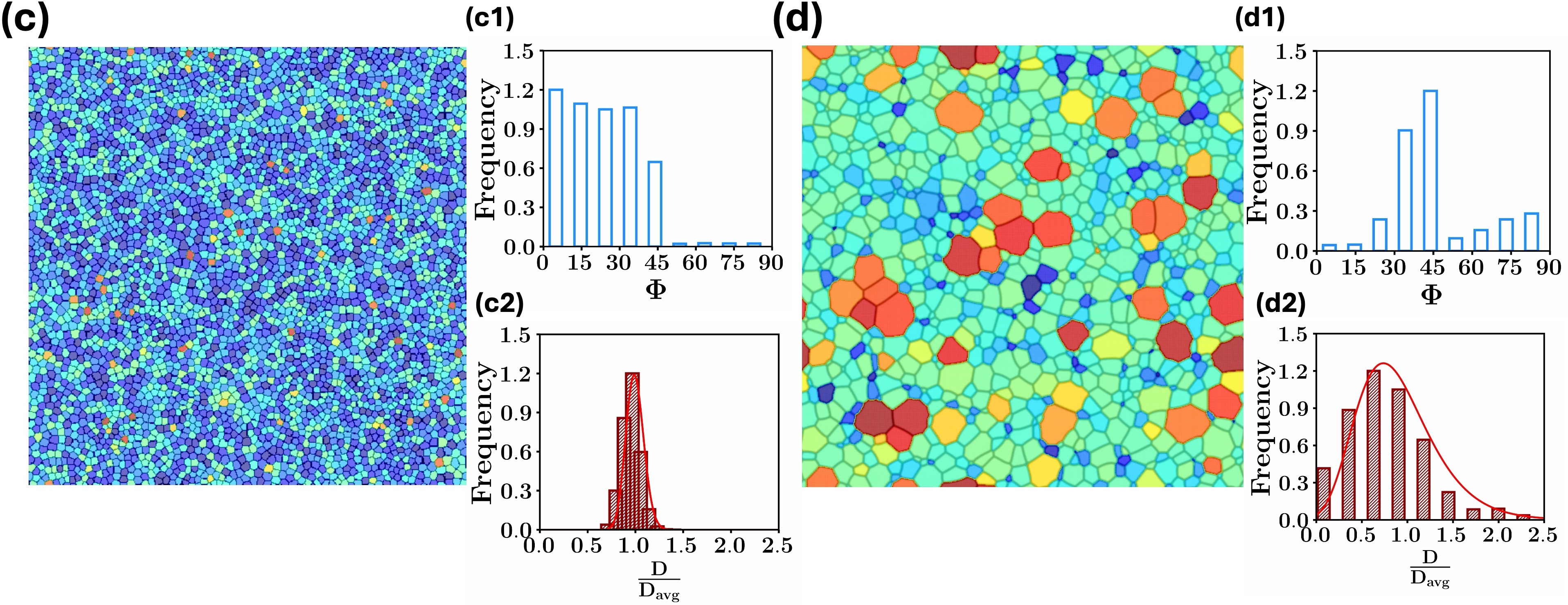}
    \end{subfigure}%

    \begin{subfigure}[t]{\textwidth}
        \centering
        \includegraphics[width=0.75\linewidth]{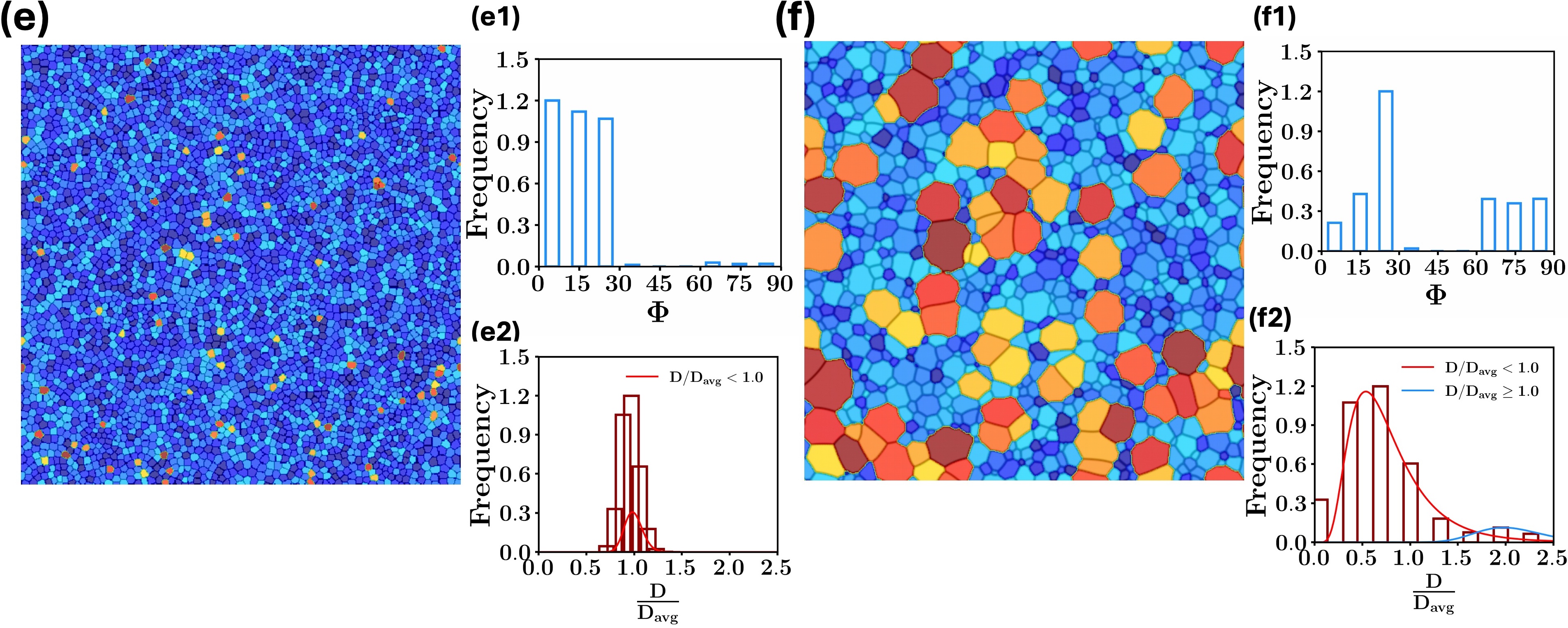}
    \end{subfigure}%

    \begin{subfigure}[t]{\textwidth}
        \centering
        \includegraphics[width=0.75\linewidth]{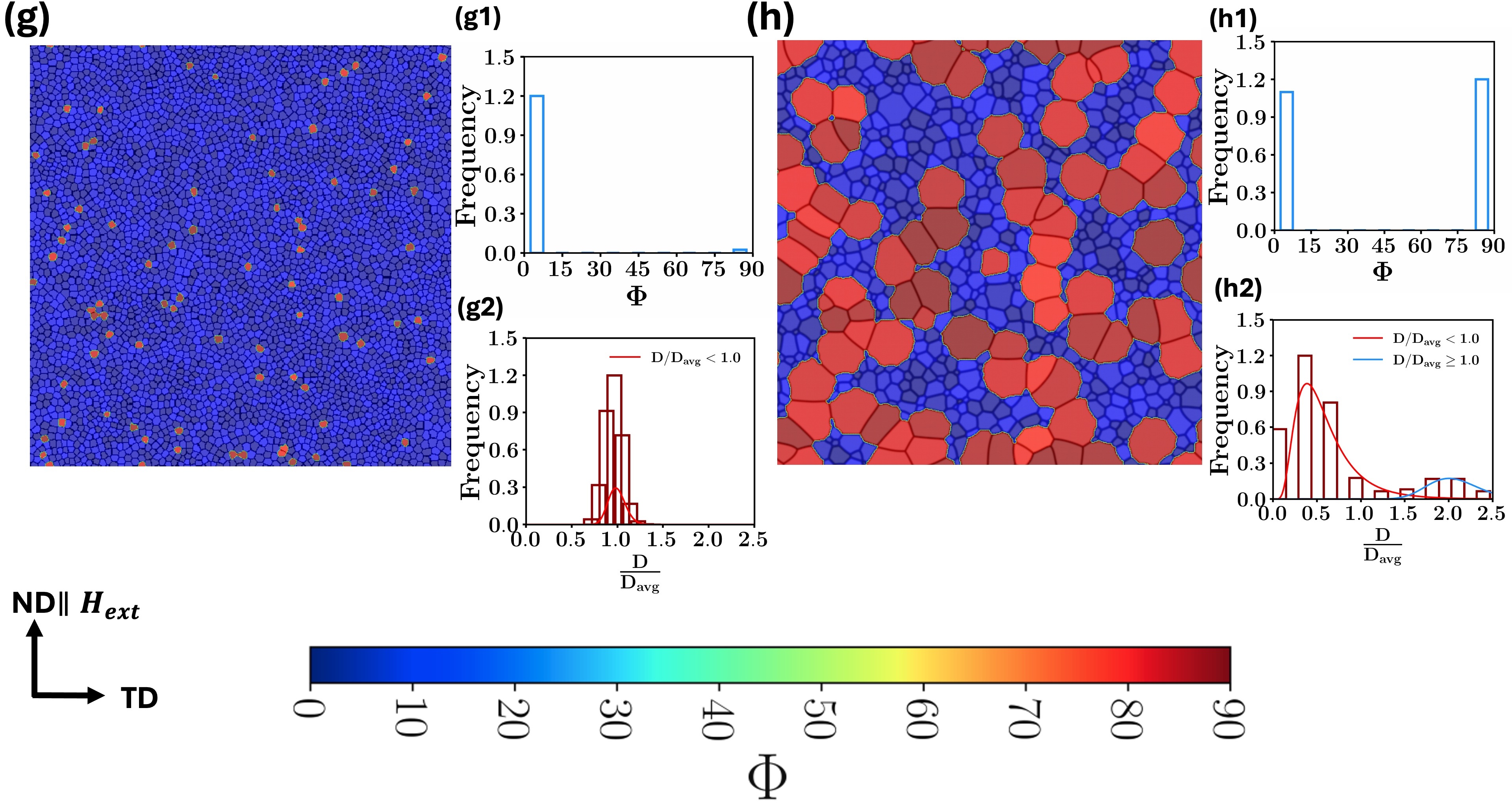}
    \end{subfigure}%
\caption{Comparison of the initial states at $t = 0$ (microstructure, orientation distribution, and size distribution) with the evolved states after $t = 5000$ across four distinct initial crystallographic textures: (a--a2) \& (b--b2) for Set 1, (c--c2) \& (d--d2) for a Set 2, (e--e2) \& (f--f2) for Set 3, and (g--g2) \& (h--h2) for Set 4. The evolved GSD show clusters of larger grains embedded within a fine grained matrix, particularly for Sets 3 and 4, indicating bimodal characteristics driven by localized magnetic orientation selection. The color bar represents the grain orientations between $(0^\circ$--$90^\circ)$.}
\label{fig:texture}
\end{figure*}

\begin{figure}[tbph]
    \centering
    \includegraphics[width=.6\linewidth]{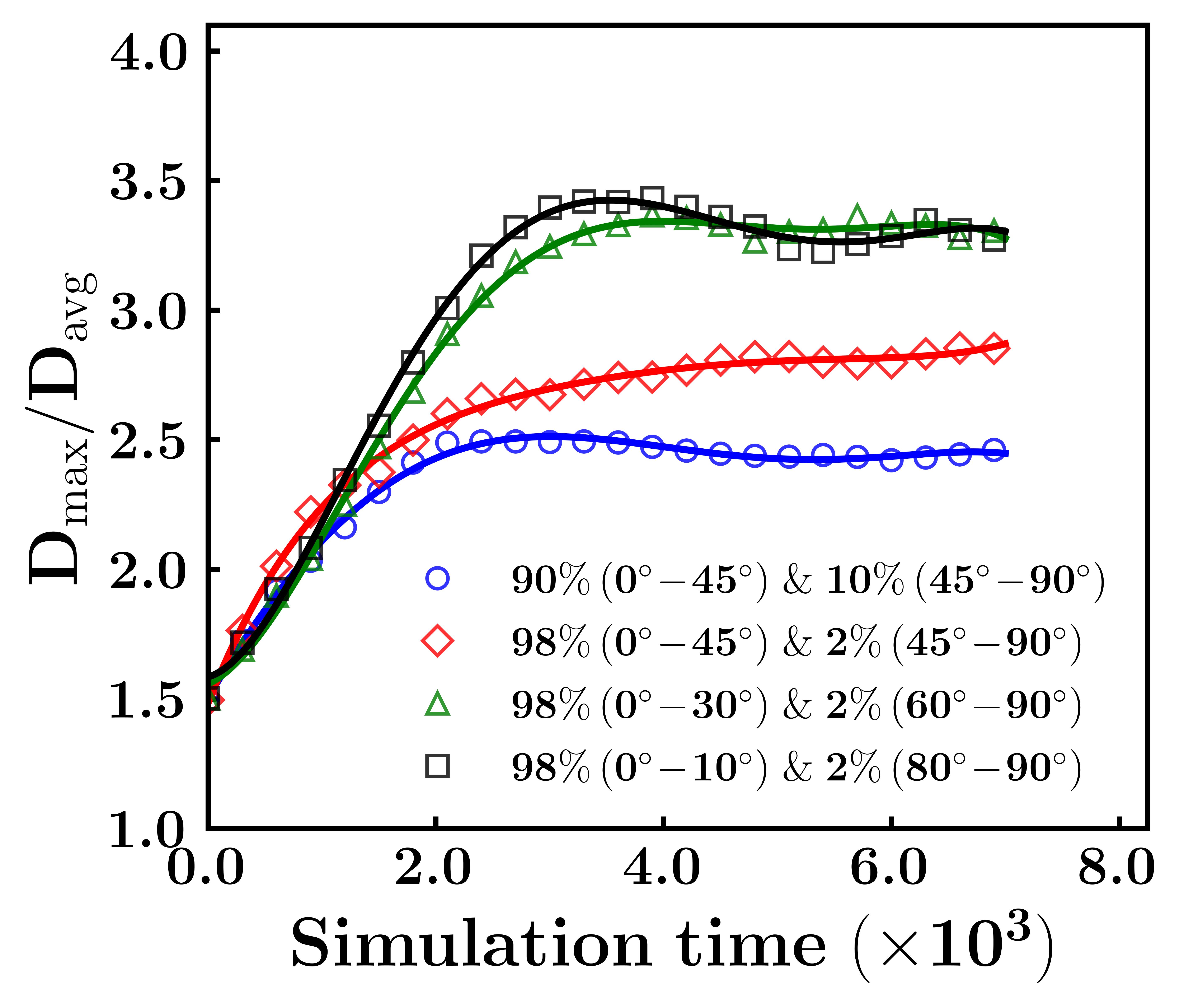} 
   \caption{
Evolution of the normalized maximum grain size ratio, $D_{\max}/D_{\mathrm{avg}}$, as a function of simulation time under $H_{\mathrm{ext}} = 8H_{\mathrm{ref}}$ field for different sets of crystallographic texture and orientation bias.}
     \label{fig-max_D_vs_D}
\end{figure}

For all the cases, we apply the magnetic field along the ND direction. Figure~\ref{fig:texture}(a) displays the initial microstructure of Set 1. Figures~\ref{fig:texture}(a1)-(a2) show the corresponding Euler angle and the normalized grain size distribution (expressed in terms of $D/D_{\text{avg}}$). Figure~\ref{fig:texture}(b) displays the evolved microstructure after 5000 simulation time. As expected, we observe preferential growth of the grains oriented within $45^\circ$-$90^\circ$ as clearly demonstrated by the corresponding Euler angle distribution in Fig.~\ref{fig:texture}(b1). However, despite this magnetic orientation selection, the growth remains relatively homogeneous. The grain size distribution broadens but does not develop a pronounced long tail toward very large grains (Fig.~\ref{fig:texture}(b2)). This is also confirmed by the temporal evolution of the $D_{\max}/D_{\text{avg}}$ ratio (marked by the blue curve in Fig.~\ref{fig-max_D_vs_D}) increases slowly with time; exhibiting the normal growth kinetics~\cite{humphreys1995recrystallization, patterson2026transition} . 

Figures~\ref{fig:texture}(c),(e),(g) show the initial microstructures for Sets 2-4, respectively, while Figs.~\ref{fig:texture}(c1),(e1),(g1), show their corresponding initial orientation distributions. The initial grain size distributions, shown in Figs~\ref{fig:texture}(c2),(e2),(g2) are relatively narrow and centered around $D_{\max}/D_{\text{avg}}\approx 1$, indicating that the initial microstructures are not strongly biased. Therefore, any subsequent development of a broad or bimodal grain size distribution can be attributed primarily to the coupling between the imposed orientation bias and the magnetic driving force.

The evolved microstructures for Sets 2–4 after 5000 simulation steps are shown in Figs~\ref{fig:texture}(d),(f),(h), respectively. We observe an interesting trend. In Set 2, the magnetically favored grains are initially rare but distributed over the broad range $(45^\circ$--$90^\circ)$. These grains grow preferentially at the expense of the surrounding lower $\Phi$ matrix, producing a broader grain size distribution than in Set 1. In Sets 3 and 4, the orientation separation between the major and minor populated grains is further increased. Consequently, the selective growth of the high $\Phi$ grains becomes more pronounced, leading to a distinct bimodal nature due to the emergence of a limited number of significantly enlarged grains embedded within a fine-grained matrix. Furthermore, a direct comparison between these sets clearly demonstrates that a stronger orientation bias and decrease in the orientation spread significantly amplifies selective growth, leading to the formation of significantly larger grains and a stronger crystallographic texture (Figs.~\ref{fig:texture}(f),(h)).

This trend is evident in both the orientation and grain size distributions. The evolved orientation distributions (See Figs.~\ref{fig:texture}(d1),(f1),(h1)) show an increase of high $\Phi$ grains, confirming the development of a stronger magnetic texture. At the same time, the corresponding normalized grain size distributions in Figs.~\ref{fig:texture}(d2),(f2),(h2) become progressively broader and increasingly right skewed, with long tails extending toward large values of $D_{\max}/D_{\text{avg}}$. This behavior is characteristic of bimodal grain growth, where a small number of favorably oriented grains grow relative to the surrounding matrix.

This bimodal behavior is further confirmed by the temporal evolution of the $D_{\max}/D_{\text{avg}}$ as shown in Fig.~\ref{fig-max_D_vs_D}. For Set 1, $D_{\max}/D_{\text{avg}}$ remains below approximately 2.5 throughout the simulation, consistent with normal grain growth. Contrarily, Sets 2-4 exceed this threshold, reaching approximately 2.8 for Set 2 (red curve), 3 for Set 3 (green curve), and 3.3 for Set 4 (black curve), respectively. This increase in $D_{\max}/D_{\text{avg}}$ becomes more pronounced as the initial orientation distribution narrows around low $\Phi$ matrix grains while a small fraction of high $\Phi$ grains is retained. This demonstrates that bimodal growth behavior is not controlled solely by the presence of magnetically favored grains, but by the combined effect of their scarcity, their orientation contrast with the matrix, and the magnetic driving force.

\subsubsection*{Case 3: Dynamic control through removal and reversal of the magnetic field}
After demonstrating that an applied magnetic field can selectively promote the growth of favorably oriented grains, we next investigate whether the evolving texture can be dynamically controlled by changing the magnetic-field condition during GG. This case is designed to examine whether the magnetic field acts only as an initial selection bias or whether continued field application is required to sustain the preferential growth of specific orientations.

\begin{figure}[tpbh]
\centering
\includegraphics[width=0.9\linewidth]{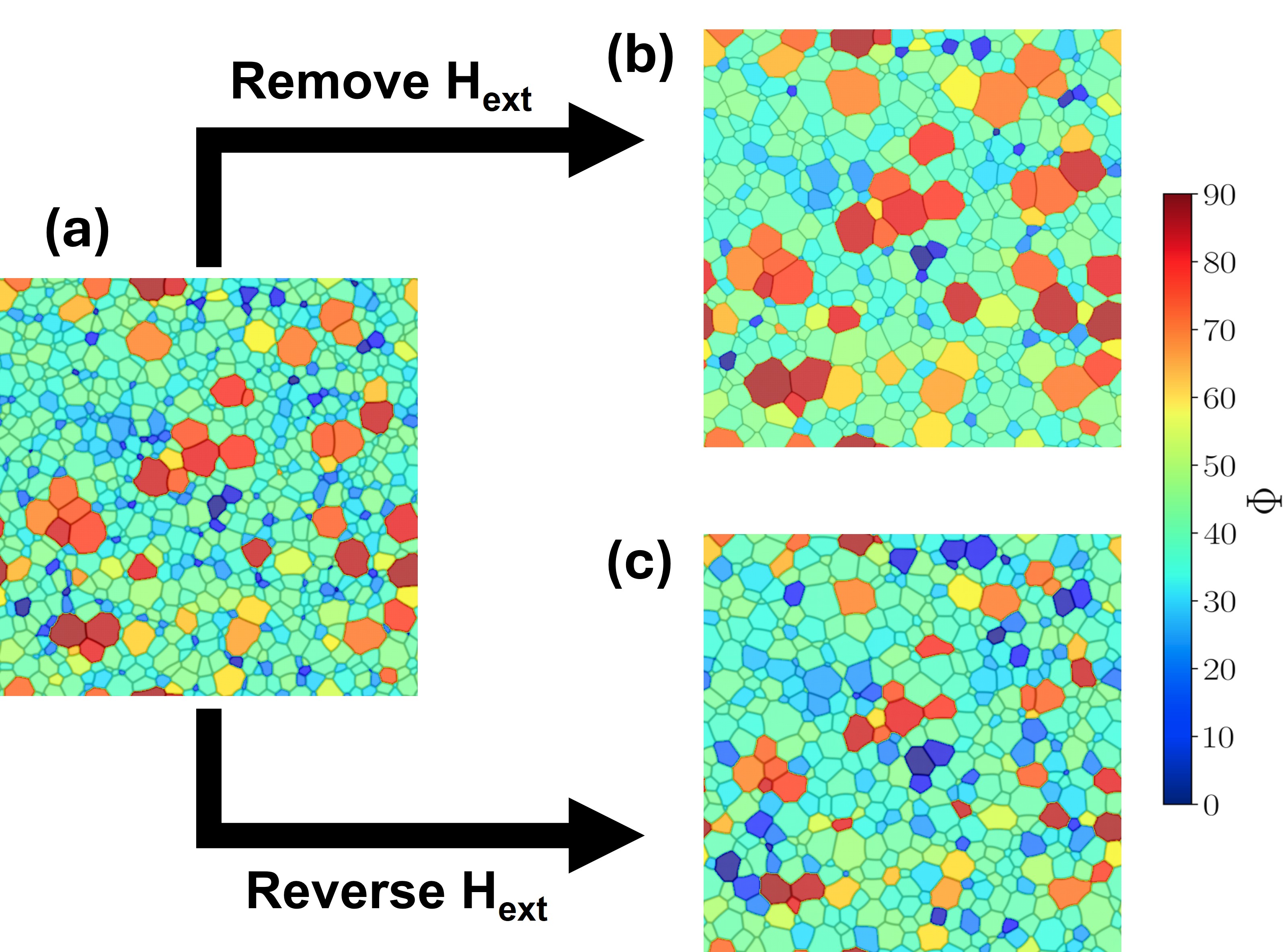} 
\caption{Microstructural evolution under removed and reversed magnetic field conditions. (a) Shows the microstructure at simulation time $t = 5000$, obtained under a magnetic field applied along the normal direction (ND), which serves as the initial configuration for subsequent evolution. Evolved morphology after an additional 5000 simulation time: (b) after complete removal of the magnetic field, and (c) complete reversal of the applied field, applied along the TD. Removal of the field eliminates the orientation selection induced by the magnetic field and leads to normal coarsening behavior, whereas reversal of the field along TD promotes the growth of a different subset of grains that were not previously favored.}
\label{fig-reverse_dir_5000_8href}
\end{figure}

To examine that, we use the evolved Set 2 microstructure after 5000 simulation time as the starting point. This microstructure, shown in Fig.~\ref{fig-reverse_dir_5000_8href}(a), is obtained under a magnetic field applied along the ND direction and already contains a magnetically biased grain structure, where favorably oriented grains have undergone preferential growth.

Starting from this common microstructure, we investigate two distinct field-control routes. In the first case, the external magnetic field is completely removed, and the system is allowed to evolve further without a magnetic driving force. The resulting microstructure is shown in Fig.~\ref{fig-reverse_dir_5000_8href}(b). The field removal eliminates the orientation-selective magnetic driving force, and the subsequent microstructure evolution is primarily governed by curvature-driven GG. As a result, the microstructure continues to coarsen, but without a strong additional preference for any particular crystallographic orientation. Importantly, the grains that had already grown during the initial magnetic field treatment remain comparatively large, indicating that the texture developed during the first stage is not immediately erased upon removal of the field.

In the second case, instead of removing the field, we change the magnetic field direction from ND to TD. The corresponding evolved microstructure is shown in Fig.~\ref{fig-reverse_dir_5000_8href}(c). Now, reorienting the field changes the crystallographic orientations that are magnetically favored. This reorientation reduces the growth advantage of the previously favored grains while a different subset of grains begins to experience preferential magnetic driving force. Compared with the previous case, this case shows a more pronounced redistribution of growth, indicating that the grain growth pathway remains sensitive to the direction of the applied field even after a biased texture has already developed.

\begin{figure}[htbp]
\centering
\includegraphics[width=1.05\linewidth]{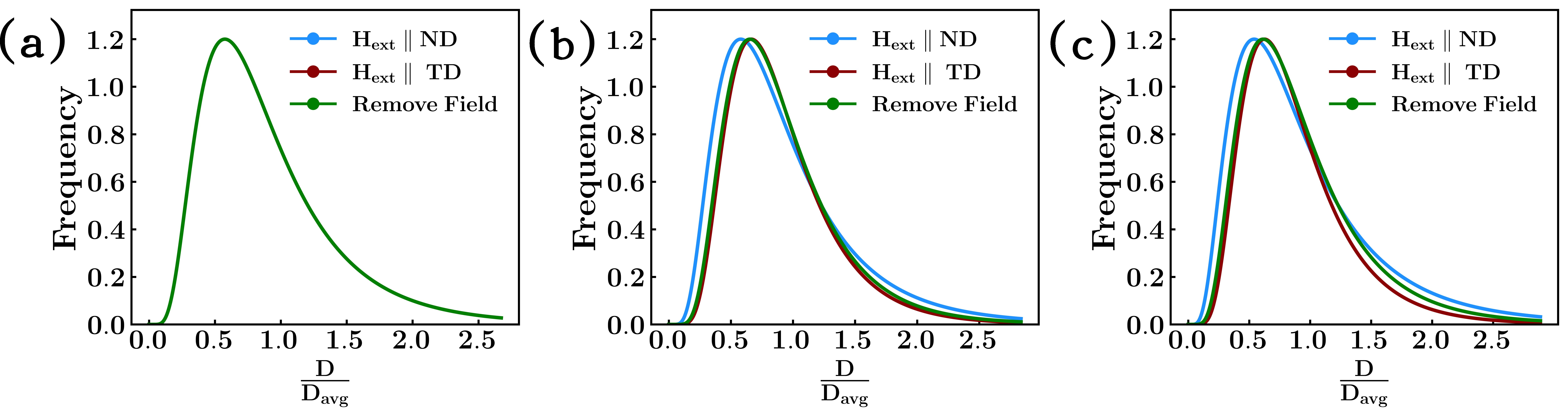} 
\caption{Evolution of grain size distributions under three different conditions: continuous application of the magnetic field along the normal direction (ND), reversal of the field to the transverse direction (TD), and complete removal of the field after $t = 5000$. Results are shown at (a) $t = 5000$, (b) $t = 8000$, and (c) $t = 10000$.  While field removal leads to conventional grain coarsening, field reversal promotes the growth of grains favored by the new field orientation, modifying grain size distribution.}
\label{fig-dist_reverse_t_5000}
\end{figure}

To further quantify the effect of dynamic field control, we compare the grain size distributions for three field histories: (a) continued application of the magnetic field along ND, (b) removal of the magnetic field, and (c) reorientation of the magnetic field along TD. For all the cases, we use the evolved Set 2 microstructure after 5000 simulation steps under the ND directed field as the common starting point. Therefore, the distributions at the initial stage of this comparison are identical, as shown in Fig.~\ref{fig-dist_reverse_t_5000}(a).

After an additional 3000 simulation time steps, differences begin to emerge among the three field histories, as shown in Fig.~\ref{fig-dist_reverse_t_5000}(b). The case with continued field application along ND shows a slightly broader distribution and a more pronounced right tail (blue curve). This indicates continued preferential growth of the grains that were favored during the initial field treatment. In contrast, the reversed-field case, where the field is reoriented along TD, shows a relatively reduced large-grain tail (brown curve). This suggests that changing the field direction partially suppresses the growth advantage of the previously favored grains and redirects the magnetic driving force toward the grains favored under the new field orientation. The field-removed case evolves mainly through curvature-driven coarsening and therefore remains close to the initial biased state, without strongly amplifying the large-grain tail.

At the later stage, Fig.~\ref{fig-dist_reverse_t_5000}(c), the same trend is retained. The ND field case maintains the broadest distribution and the most extended right tail, consistent with sustained orientation-selective growth. The reversed-field case shows a comparatively narrower distribution and a leftward-shifted tail, indicating that field reversal counteracts or redistributes the selective growth established during the initial ND-field stage. The field removal case remains intermediate, reflecting continued coarsening after the magnetic driving force is removed while still preserving some of the grain-size heterogeneity inherited during the initial field treatment. 

\section{Discussion} \label{sec:discussion}
In this study, we focus on delineating how directional magnetic fields influence GB migration, texture evolution, and the bimodal growth characteristics in non-magnetic polycrystals. Our systematic investigation reveals that the magnetic field does not uniformly accelerate grain growth. Rather, the external magnetic field acts as an orientation-selective driving force whose effect depends on the crystallographic orientation distribution, the field direction, and its history. 

We first establish the orientation-dependent nature of magnetic field-driven boundary migration in a bicrystal system (Fig.~\ref{fig:contour}), and establish the basic selection mechanism: magnetic field-driven GB migration occurs only when neighboring grains have different magnetic free energy densities. This difference arises from the orientation dependence of the susceptibility tensor. Since Bi has a hexagonal crystal structure, its magnetic susceptibility is anisotropic, with different susceptibilities along the basal and c-axis directions. As a result, rotating the crystallographic c-axis relative to the applied field changes the effective susceptibility of a grain and, consequently, its magnetic free energy density~\cite{bandyopadhyay2025grain, liang2022phase}. Therefore, the field acts as a crystallographic selector rather than as a uniform coarsening accelerator. The increase in migration rate with misorientation (see Fig.~\ref{fig:interface_displacement}) confirms that the magnetic driving force scales with the susceptibility contrast between neighboring grains.

In a bicrystal with a planar boundary, curvature effects are non-existent, and the magnetic free energy difference is the only driving force. However, in a polycrystalline microstructure, curved boundaries, triple junctions, and local grain-size variations can drive capillarity-driven migration even in the absence of a magnetic field. Therefore, the competition between the magnetic driving force (arising from orientation-dependent free-energy differences) and the capillary driving force (associated with boundary curvature) determines overall grain growth in a polycrystal, motivating us to extend the analysis to polycrystalline systems. 

We first study the effect of the external magnetic field on grain growth in an initially unbiased polycrystal (Fig.~\ref{fig-micropoly}), showing the competition between capillarity and magnetic free-energy contributions. At lower fields, the magnetic term is not sufficient to alter the dominant curvature-driven coarsening pathway. However, once the field exceeds a threshold, the magnetic contribution becomes kinetically relevant, as reflected by the increase in the growth rate constant (Fig.~\ref{magnetic_to_capillary_ration1}). However, this enhanced kinetic response is not accompanied by a pronounced change in the grain size distribution (Fig.~\ref{fig-microstructure8href} and ~\ref{figmax_D_vs_D_nt_td_without_field}). Instead, the dominant effect is the development of crystallographic texture. 

Additionally, the polycrystalline simulations in Fig.~\ref{fig:texture} demonstrate that magnetic orientation selection becomes a collective statistical effect in polycrystals. When the favored orientations are abundant, magnetic selection is distributed across a larger number of grains, and the system undergoes a normal GG (Fig.~\ref{fig:texture}(a)-(b2). As a result, the microstructure coarsens with an orientation bias, but without developing the strong size disparity: a classic characteristic of bimodal GG. However, when the favored orientations are strongly separated from the matrix orientation distribution, the same magnetic driving force becomes localized onto a small subset of grains (Figs.~\ref{fig:texture}(c)-(h2)). This localization of driving force promotes selective coarsening, leading to the emergence of oversized grains, a broadened right-skewed grain size distribution (Figs.~\ref{fig:texture}(d2),(f2),(h2)), with $D_{\max}/D_{\mathrm{avg}}>2.5$ resulting in the onset of bimodal behavior (Fig~\ref{fig-max_D_vs_D}). This reflects a transition from distributed to localized magnetic selection.

The dynamic field cases demonstrate that magnetic selection is history dependent (Fig.~\ref{fig-reverse_dir_5000_8href}). Continued application of the original field reinforces the initially selected grain population, while removal of the field suppresses further magnetic amplification and normal curvature-driven coarsening. Field reorientation modifies the selection pathway by changing which orientations are favored (Fig.~\ref{fig-dist_reverse_t_5000}). Thus, the final microstructure depends not only on field magnitude and initial texture but also on the sequence of field application, indicating that magnetic fields serve as dynamic control parameters rather than merely static external driving forces. This progression from bicrystal migration to polycrystalline growth and dynamic field control provides a mechanistic framework for understanding how magnetic orientation selection can be used to tailor GG pathways in non-magnetic materials.


Currently, the study is limited to 2D; since real polycrystalline GBs are governed by 3D boundary curvature and junction topology, performing 3D simulations is a natural extension of this work to test the robustness of the magnetic orientation selection mechanism under realistic microstructural conditions. Additionally, the current model assumes isotropic GB energy and mobility; incorporating anisotropic boundary properties would allow exploration of the rigorous interplay among the magnetic contribution, other interfacial effects, and their contributions to the bimodal nature of grain growth. Finally, additional dynamic field protocols, including field rotation, periodic reversal, and pulsed field application, should be investigated to determine whether magnetic fields can accelerate grain growth while limiting unwanted texture development.

Despite our simplifying assumptions, the present model provides a robust framework for isolating magnetic orientation selection as an independent mechanism for bimodal characteristics. The emergence of selective coarsening without particle pinning, GB energy anisotropy, or imposed mobility contrast indicates that directional magnetic fields alone can bias grain populations and regulate the bimodal growth characteristics in non-magnetic polycrystals.

Taken together, our findings establish directional magnetic fields as an independent thermodynamic control variable for GG in non-magnetic polycrystals. The governing mechanism is not a uniform enhancement of GB mobility; rather, it is an orientation-dependent magnetic free-energy bias that selectively stabilizes and amplifies grains with favorable crystallographic alignment. By changing the magnitude, direction, and temporal history of the applied field, this bias can be used to regulate the transition between normal coarsening, texture development, and bimodal grain growth. Thus, the magnetic field history provides a physically grounded route for steering microstructural evolution beyond conventional curvature-driven GG.

\section{Conclusion} \label{sec:conclusion}
This study shows that directional magnetic fields can drive orientation-selective grain growth in non-magnetic polycrystals through susceptibility anisotropy. Bicrystal simulations first establish that grain boundary migration arises from the magnetic free-energy difference between neighboring grains, confirming that the field assists the selection of the preferred crystallographic orientation rather than uniformly accelerating grain coarsening.

Studies on texture-informed polycrystalline systems demonstrate that this selection mechanism leads to different growth regimes depending on the initial orientation distribution. When magnetically favored orientations are abundant, the system exhibits biased but largely normal coarsening. However, when favored grains are rare and exhibit a large magnetic free-energy difference relative to the matrix, magnetic selection becomes localized, producing bimodal Growth behavior. The study further shows that bimodal coarsening can emerge without particle pinning, GB energy anisotropy, or imposed mobility contrast. Instead, the bimodal characteristics of g are driven by the coupling between the initial orientation bias and the magnetic free-energy contrast.

Dynamic field simulations demonstrate that magnetic selection is history dependent. Continued field application reinforces the selected grain population, field removal suppresses further magnetic amplification, and field reorientation modifies the subsequent growth pathway. Overall, directional magnetic fields provide an independent thermodynamic route to control grain growth, texture evolution, and bimodal coarsening in non-magnetic polycrystals.

\appendix
\section{Summary of dimensionless parameters used in this study}
\label{appendix_dimensionless}

All parameters in this study are expressed in their scaled, dimensionless form. This section describes the procedure used to non-dimensionalize the material model parameters applied in the simulations. To achieve this, we define a characteristic length ($L_c$), characteristic energy ($E_c$), and characteristic time ($\tau_c$), which serve as the basis for scaling all parameters\cite{bandyopadhyay2025grain}. We considered the grain boundary energy of the system to be $\sigma_{GB} = 0.155~\mathrm{J/m^2}$ \cite{ocak2008interfacial} and the interfacial width $l_{GB} = 1.6~\mathrm{nm}$. During the simulations, a minimum of five grid points was always maintained across the diffuse interface to ensure adequate spatial resolution. Accordingly, the grid spacing was chosen as $\delta  = \frac{l_{GB}}{5} = 0.32~\mathrm{nm}.$

To obtain the characteristic energy scale ($E_c$), it was estimated using the Boltzmann constant $k_B = 1.38 \times 10^{-23}~\mathrm{J\,K^{-1}}$ and the operating temperature $T = 225^\circ\mathrm{C}$. Thus, the characteristic energy is given by
\begin{equation}
E_c = k_B T = 7.2864 \times 10^{-21}~\mathrm{J}
\label{eq:characteristic energy}
\end{equation}
Additionally, the characteristic length scale ($L_c$) was determined from the ratio between the characteristic energy and the grain boundary energy as
\begin{equation}
L_c = \sqrt{\frac{E_c}{\sigma_{GB}}}
     = 2.1682 \times 10^{-10}~\mathrm{m}
\label{eq:characteristic length}
\end{equation}
To estimate the characteristic time scale ($\tau_c$), the relaxation parameter $L$ employed in the phase-field simulations was considered as \cite{liang2022phase}
\begin{equation}
L = \frac{4m}{3l_{GB}}
\label{eq:relaxation parameter}
\end{equation}
where the grain boundary mobility $m$ is expressed as
\begin{equation}
m = m_0 \exp\left(-\frac{Q}{k_B T}\right).
\label{eq:pre-exponent}
\end{equation}
Here, $m_0 = 1.1 \times 10^{24}~\mathrm{m^4\,J^{-1}\,s^{-1}}$ represents the pre-exponential factor, and $Q = 3.38~\mathrm{eV}$ denotes the activation energy \cite{molodov1998true}. Using these parameters, the relaxation coefficient was evaluated as $L = 5.3539~\mathrm{m^3\,J^{-1}\,s^{-1}}$.

Accordingly, the characteristic time scale was obtained as 
\begin{equation}
\tau_c = \frac{L_c^3}{L E_c}
       = 2.6129 \times 10^{-10}~\mathrm{s}
\label{eq:characteristic time}
\end{equation}
The simulation-specific material parameters were subsequently determined using the interfacial properties of the system. The bulk free-energy coefficient $B$ (Double-well potential) was calculated as a function of the grain boundary energy $\sigma_{GB}$ and interfacial width $l_{GB}$ according to
\begin{equation}
B = \frac{6\sigma_{GB}}{l_{GB}}
  = 5.8125 \times 10^{8}~\mathrm{J\,m^{-3}}
\label{eq:double-well potential}
\end{equation}
Similarly, the gradient energy coefficient associated with the grain order parameters was estimated as \cite{liang2022phase}
\begin{equation}
\kappa_{\eta} = \frac{3\sigma_{GB} l_{GB}}{4}
              = 1.86 \times 10^{-10}~\mathrm{J\,m^{-1}}
\label{eq:kappa_eta}
\end{equation}

\section{Average Grain Size Estimation and Driving Force Analysis}
\label{appn}
To calculate the average grain size, a two-dimensional polycrystalline system was employed, in which the average grain size was estimated analytically using a stereological method~\cite{astm2004standard,1991214}. The simulation domain consists of a square system of dimensions $L_x \times L_y$ containing approximately $N$ grains. Thus, the total system area is given by
\begin{equation}
A = L_x L_y 
\end{equation}
Assuming nearly equiaxed and randomly distributed grains, the average grain area can be approximated as
\begin{equation}
A_{\mathrm{avg}} = \frac{A}{N}
\label{eq: area}
\end{equation}
In a two-dimensional microstructure, each grain can be represented as an equivalent circular grain. Therefore, one can represent the average area $A_{avg}$ as:
\begin{equation}
A_{\mathrm{avg}}
=
\pi
\left(
\frac{D_{\mathrm{avg}}}{2}
\right)^2
\label{eq:average area}
\end{equation}
which yields the analytical expression for the average grain diameter:
\begin{equation}
D_{\mathrm{avg}}
=
\sqrt{
\frac{4L_xL_y}{\pi N}
}
\label{eq:mean diameter}
\end{equation}
The curvature-induced interfacial energy \cite{molodov2007annealing,porter2009phase} associated with grain growth can then be expressed as 
\begin{equation}
p_{m}^{\mathrm{interface}}
=
\frac{2\sigma_{GB}}{D_{\mathrm{avg}}}
\label{eq:curvature driving force}
\end{equation}
where $\sigma_{GB}$ denotes the grain boundary energy.

In the absence of an external magnetic field, GG is primarily governed by the reduction of total interfacial energy. Consequently, GBs migrate toward their centers of curvature, resulting in progressive grain coarsening and an increase in the average grain size.

Under an applied magnetic field, however, an additional orientation-dependent driving force emerges due to the magnetic free-energy difference between neighboring grains. We can then rewrite the magnetic driving force \cite{molodov2007annealing,molodov2014grain} acting on a grain boundary as: 
\begin{equation}
p_{m}^{\mathrm{magnetic}}
=
\frac{\mu_0 \Delta \chi H^2}{2}
\left(
\cos^2\theta
-
\frac{1}{n}
\sum_{j=1}^{n}
\cos^2\theta_j
\right)
\label{eq:magnetic driving force}
\end{equation}
where $\mu_0$ is the permeability of free space, $\Delta \chi$ is difference of the magnetic susceptibility anisotropy of $\chi_{\parallel} $ and  $\chi_{\perp}$, $H$ is the applied magnetic field strength, and $\theta$ and $\theta_j$ represent the orientations of the principle axis of considered grain and its neighboring grains relative to the applied magnetic field direction.

In a polycrystalline system, grain boundary migration is therefore governed by the competition between curvature-driven capillary forces and magnetic anisotropy-driven forces. To quantify their relative influence, the ratio between magnetic and interfacial driving forces may be expressed as
\begin{equation}
\Gamma
=
\frac{
p_{m}^{\mathrm{magnetic}}
}{
p_{m}^{\mathrm{interface}}
}
=
\frac{
D_{\mathrm{avg}}
}{
2\sigma_{GB}
}
\cdot
\frac{
\mu_0 \Delta \chi H^2
}{2}
\left(
\cos^2\theta
-
\frac{1}{n}
\sum_{j=1}^{n}
\cos^2\theta_j
\right)
\label{eq:magnetic to capillary ratio}
\end{equation}
This dimensionless parameter provides a direct measure of the competition between magnetic and capillary effects during microstructural evolution. When $\Gamma \ll 1$, grain growth remains predominantly driven by curvature. In contrast, as $\Gamma$ increases, the magnetic driving force becomes increasingly significant, leading to orientation-selective growth and magnetic-field-assisted grain growth.

\bibliographystyle{ieeetr}
\bibliography{references}

\end{document}